\documentclass{aa}

\usepackage{graphicx}
\usepackage{amsmath}
\usepackage{amssymb}
\usepackage{xspace}
\usepackage{commath}
\usepackage{times}
\usepackage{bm} 
\usepackage{balance}
\usepackage{hyperref}
\usepackage{mathtools}
\usepackage{stmaryrd}
\usepackage{tabulary}
\usepackage{xcolor}
\usepackage{graphicx}
\usepackage{txfonts}
\usepackage{orcidlink}

\usepackage{afterpage}
\usepackage{multicol}
\newsavebox{\shortpagebox}

\makeatletter
\newcommand{\shortpage}[1]{\par
  \setbox\shortpagebox=\vbox{\strut #1\par}\afterpage{\onecolumn
    \begin{multicols}{2}
    \unvbox\AP@partial
    \end{multicols}}\unvbox\shortpagebox
\par}
\makeatother

\hypersetup{dvips, colorlinks=true, linkcolor=blue, citecolor=blue, filecolor=blue, urlcolor=blue}

\defcitealias{Tep2022}{T22}

\usepackage{acronym}
\newacro{RR}{resonant relaxation}
\newcommand{\RR}{\ac{RR}}

\newacro{NR}{non-resonant relaxation}
\newcommand{\NR}{\ac{NR}}

\newacro{VRR}{vector resonant relaxation}
\newcommand{\VRR}{\ac{VRR}}

\newacro{DF}{distribution function}
\newacroplural{DF}[DFs]{distribution functions}
\newcommand{\DF}{\ac{DF}}
\newcommand{\DFs}{\acp{DF}}

\newacro{BH}{black hole}
\newacroplural{BH}[BHs]{black holes}
\newcommand{\BH}{\ac{BH}}
\newcommand{\BHs}{\acp{BH}}

\newacro{LBD}{Lynden-Bell daemon}
\newcommand{\LBD}{\ac{LBD}}

\newacro{FP}{Fokker--Planck}
\newcommand{\FP}{\ac{FP}}

\newcommand{\re}{\mathrm{e}}
\newcommand{\rt}{\mathrm{t}}
\newcommand{\rr}{\mathrm{r}}
\newcommand{\rc}{\mathrm{c}}

\newcommand{\ra}{\mathrm{a}}
\newcommand{\rp}{\mathrm{p}}

\newcommand{\rd}{{\mathrm{d}}}

\newcommand{\rHU}{{\mathrm{HU}}}
\newcommand{\Rv}{R_{\mathrm{v}}}
\newcommand{\sgn}{\mathrm{sgn}}

\newcommand{\tlast}{t_{\mathrm{last}}}
\newcommand{\Nrun}{N_{\mathrm{run}}}

\newcommand{\p}{\partial}

\newcommand{\rra}{r_{\ra}}
\newcommand{\rrp}{r_{\rp}}
\newcommand{\vr}{v_{\rr}}
\newcommand{\vt}{v_{\rt}}

\newcommand{\bL}{\boldsymbol{L}}

\newcommand{\mbF}{\boldsymbol{\mathcal{F}}}
\newcommand{\mF}{\mathcal{F}}

\newcommand{\Lo}{L_{0}}

\newcommand{\Jr}{J_{\rr}}

\newcommand{\Ftot}{F_{\mathrm{tot}}}
\newcommand{\Frot}{F_{\mathrm{rot}}}

\newcommand{\bx}{\boldsymbol{x}}
\newcommand{\bv}{\boldsymbol{v}}

\newcommand{\bw}{\boldsymbol{w}}

\newcommand{\bJ}{\boldsymbol{J}}
\newcommand{\bJc}{\boldsymbol{J}_{\rc}}

\newcommand{\br}{\boldsymbol{r}}
\newcommand{\hr}{\hat{\br}}
\newcommand{\bD}{\boldsymbol{D}}

\newcommand{\dvPar}{\langle\Delta v_{\parallel} \rangle}
\newcommand{\dvParLoc}{ \Delta v_{\parallel}}
\newcommand{\dvPerpLoc}{\Delta v_{\perp}}

\newcommand{\dE}{\langle \Delta E \rangle}
\newcommand{\dL}{\langle \Delta L \rangle}
\newcommand{\dESq}{\langle (\Delta E)^2 \rangle}
\newcommand{\dLSq}{\langle (\Delta L^2) \rangle}
\newcommand{\dEL}{\langle \Delta E \Delta L \rangle}

\newcommand{\dLz}{\langle \Delta \Lz \rangle}
\newcommand{\dLzSq}{\langle (\Delta \Lz^2 \rangle}
\newcommand{\dELz}{\langle \Delta E \Delta \Lz \rangle}
\newcommand{\dLLz}{\langle \Delta L \Delta \Lz \rangle}

\newcommand{\dvParSq}{\langle (\Delta v_{\parallel})^2 \rangle}
\newcommand{\dvPerpSq}{\langle (\Delta v_{\perp})^2 \rangle}

\newcommand{\half}{\tfrac{1}{2}}

\newcommand{\Lz}{L_{z}}
\newcommand{\trh}{t_{\mathrm{rh}}}
\newcommand{\rh}{r_{\mathrm{h}}}

\newcommand{\cosI}{\cos \! I}
\newcommand{\erf}{\mathrm{erf}}

\newcommand{\bI}{\mathbf{I}}
\newcommand{\mbI}{\boldsymbol{\mathcal{I}}}
\newcommand{\rT}{\mathrm{T}}
\newcommand{\by}{\mathbf{y}}

\newcommand{\hbv}{\hat{\boldsymbol{v}}}

\usepackage{soul} \usepackage{xcolor}
\definecolor{aquamarine}{rgb}{0.5, 1.0, 0.83}

\begin{document}

 \title{Non-resonant relaxation of rotating globular clusters}
   \titlerunning{Non-resonant relaxation}
   
 \authorrunning{K. Tep, J.-B. Fouvry \& C. Pichon}
 
  \author{
          Kerwann~Tep\inst{1,2}\orcidlink{0009-0002-8012-4048} \thanks{E-mail: \href{mailto:tep@iap.fr}{tep@iap.fr}}\and
          Jean-Baptiste~Fouvry\inst{1}\orcidlink{0000-0002-0030-371X}\and
          Christophe~Pichon\inst{1,3,4} \orcidlink{0000-0003-0695-6735}
          }
          
 \institute{
 		Institut d'Astrophysique de Paris, CNRS and Sorbonne Universit\'e, UMR 7095, 98 bis Boulevard Arago, F-75014 Paris, France
          \and 
                Department of Physics and Astronomy, The University of North Carolina at Chapel Hill, Chapel Hill, NC 27599, USA
        \and
             IPhT, DRF-INP, UMR 3680, CEA, L'Orme des Merisiers, B\^at 774, 91191 Gif-sur-Yvette, France
        \and
             Korea Institute for Advanced Study, 85 Hoegi-ro, Dongdaemun-gu, Seoul 02455, Republic of Korea
             }

\abstract{
The long-term relaxation of rotating, spherically symmetric globular clusters is investigated through an extension
of the orbit-averaged Chandrasekhar non-resonant formalism.
A comparison is made with the long-term evolution of the distribution function in action space,
 measured from averages of sets of $N$-body simulations up to core collapse.
The impact of rotation on in-plane relaxation is found to be weak. In addition, we observe a clear match between theoretical predictions and $N$-body measurements.
For the class of rotating models considered,
we find no strong gravo-gyro catastrophe accelerating core collapse.
Both kinetic theory and simulations predict a reshuffling of orbital inclinations
from overpopulated regions to underpopulated ones.
This trend  accelerates as the amount of rotation is increased.
Yet, for orbits closer to the rotational plane,
the non-resonant prediction does not reproduce numerical measurements.
We argue that this mismatch stems from
these orbits' coherent interactions,
which are not captured by the non-resonant formalism that only addresses local deflections.
}


\maketitle

\section{Introduction}
\label{sec:intro}

Rotation is ubiquitous in stellar systems.
In effect, it provides a source of free energy,
allowing clusters to efficiently reshuffle their orbital structure
towards more likely configurations.
Yet, historically, the study of globular clusters has been mainly focused on isotropic, non-rotating, old globular clusters  
\citep{Aarseth1974,Spitzer1975,Cohn1979,Trager1995,Miocchi2013}. 
The reasons behind such simplifications are two-fold:
(i) naturally, it is easier numerically and analytically to neglect the effect of rotation; 
(ii) spherical isotropic models -- for example, the King models \citep{King1966} or the Wilson models~\citep{Wilson1975} -- provided a satisfactory zeroth-order description of the main observed dynamical properties
of globular clusters~\citep[see, e.g.,][]{McLaughlin2005}.

The last decade has seen the extraction of new data 
 -- for instance, \textit{HST}~\citep{Bellini2017} and \textit{Gaia} DR2~\citep{Bianchini2018,Sollima2019}.
 These surveys gave the astrophysical community access to numerous and detailed observations of the internal kinematics of several globular clusters of the Milky Way~\citep{Bianchini2013, Fabricius2014,Watkins2015, Ferraro2018, Kamann2018}, as well as a quantification of the degree of velocity anisotropy~\citep{Jindal2019}. 
 Using these new data sets, the historical highly symmetric cluster models are not satisfactory anymore.

Therefore, a secular theory that describes the evolution of globular clusters
and accounts for their rotation is needed to describe their long-term evolution. In particular, the Fokker-Planck theory was used to probe the impact of rotation of the long-term evolution of such clusters, both before \citep{Einsel1999} and after \citep{Kim2002} core collapse. 
Furthermore, observations show that the angular momentum distribution measured in Galactic clusters
retain the signature of their formation process~\citep{Lanzoni2018}.
While $N$-body simulations are able to reproduce these results~\citep{Tiongco2016, Tiongco2022},
the historical context and the complexity of the problem have led to few analytical explorations \citep{Geyer1983, White1987, Kontizas1989}. These remain scarce even now~\citep[see, e.g.,][]{Stetson2019,Rozier2019,Livernois2022}.

In isolated systems, a non-zero total angular momentum,
that is, the presence of rotation,
can have a significant impact on the cluster's long-term evolution,
for instance, through the ``gravo-gyro catastrophe''~\citep{Hachisu1979,Ernst2007}.
This phenomenon has been observed in a range of rotating systems,
including gas cylinders~\citep{Inagaki1978,Hachisu1979},
gaseous discs~\citep{Hachisu1982},
flattened (quasi-spherical) star clusters~\citep{Akiyama1989,Einsel1999,Ernst2007},
and clusters with embedded \BHs\@~\citep{Fiestas2010, Kamlah2022}.
Let us however note that many of these studies used a 
rotating King model to study the impact of rotation~\citep[see, e.g.,][]{ Einsel1999,Varri2012}.
Therein, changing the amount of rotation
impacts the density profile: this makes comparisons
amongst different models less clear.

The concurrent occurrence of internal rotation and a spectrum of stellar masses
can result in the formation of an oblate core of fast rotating heavy masses~\citep{Kim2004, Tiongco2021}.
More precisely, the orbital inclinations of the heaviest stars align
with respect to one another, inducing a mass segregation in the distribution
of orbital inclinations~\citep{Szolgyen2019}.
The generally agreed explanation for this phenomenon is resonant relaxation and resonant friction~\citep{Rauch1996, Meiron2019}.
While this effect concerns globular clusters,
nuclear cluster with a dominant massive \BH\@~\citep{SzolgyenKocsis2018, Foote2020, Gruzinov2020, Magnan2022,Ginat2023} also display this spontaneous alignment.
 
The objective of this paper is to explore the impact of rotation on the long-term evolution of spherically symmetric globular clusters. To achieve this goal, we extend the orbit average analysis of anisotropic Plummer globular clusters from \cite{Tep2022}
to the case of rotating clusters.
By performing tailored $N$-body simulations, we quantify the validity of Chandrasekhar's \NR\ theory \citep{Chandrasekhar1943},
and assess the importance of the {gravo-gyro catastrophe}
as well as that of collective effects \citep[i.e., the self-amplification of the star's gravitational response, see][]{Heyvaerts2010}.

This paper is organised as follows. In section~\ref{sec:NR}, we extend the \NR\ theory to rotating clusters. In section~\ref{sec:LongTerm}, we follow the long-term evolution of a series of rotating clusters using $N$-body simulations. We study the impact of rotation on core collapse and on the distribution of orbital inclinations.
We compare the \NR\ prediction with $N$-body simulations in sections~\ref{sec:Secular_Jr} (in-plane diffusion) and~\ref{sec:Secular_cosI} (out-of-plane diffusion).
Finally, we discuss our results in section~\ref{sec:discussion}.

\section{Non-resonant relaxation}
\label{sec:NR}

We consider a self-gravitating globular cluster composed of $N$ stars of individual mass ${m\!=\!M/N}$, with $M$ the cluster's total mass.
We follow the cluster's evolution through the total \DF\ in ${(\br,\bv)}$ space, ${\Frot\!=\!\Frot(\br,\bv)}$,
with $\br$ the position and $\bv$ the velocity.
All normalisations are taken to be the same as in section~{2} of \cite{Tep2022},
hereafter \citetalias{Tep2022}.
In addition to quasi-stationarity,
we assume the cluster to be spherically symmetric -- hence, with planar unperturbed orbits -- and in rotation.
Since we are interested in orbital distortion,
it is convenient to monitor the cluster's evolution in action space, via
${\Frot \!=\! \Frot(\bJ)}$, where $\bJ = (\Jr, L, \Lz)$ are the specific action coordinates.
Here, $\Jr$ is the radial action,
$L$ the norm of the angular momentum vector
and $\Lz$ its projection along the $z$-axis.
We also introduce the orbital inclination
through ${\cos I\!=\!\Lz/L}$.
Because actions are integrals of motion, we use them
to label the cluster's orbits and 
track their deformation over time.

The long-term evolution of the cluster's \DF\
is described by the \FP\ equation \citep[see, e.g., section~{7.4} of][]{Binney2008}
\begin{align}
\label{eq:def_FP}
\frac{\p \Frot (\bJ,t)}{\p t} {} &\!=\! - \frac{\p }{\p \bJ} \!\cdot\! \mbF (\bJ) \\
&\!=\! - \frac{\p }{\p \bJ}\! \cdot \!\bigg[  \bD_{1} (\bJ) \, \Frot (\bJ)\!-\! \frac{1}{2} \frac{\p }{\p \bJ} \!\cdot\! \bigg(  \bD_{2} (\bJ) \, \Frot (\bJ)  \bigg) \bigg] ,\notag
\end{align}
with $\mbF (\bJ)$ the action space flux.
Therein, we find the first-order diffusion coefficient, $\bD_{1}$,
and the diffusion matrix, $ \bD_{2}$.
Both of them describe the distortion of orbits in action space.
The next two sections briefly describe the main steps
needed to compute this diffusion flux in equation~\eqref{eq:def_FP}.

\subsection{Local velocity deflection coefficients}
\label{subsec:loc_vel_diff_coeffs}

Let us consider a test star of mass $m$ and velocity $\bv$.
As a result of the cluster's finite number of constituents,
this test star is subject to perturbations around its mean field trajectory,
driving an irreversible diffusion of its velocity.
We call this long-term relaxation process,
sourced by successive, uncorrelated pairwise deflections,
the \NR\ relaxation.
Following appendix~{L} of \citet{Binney2008} (p.~836), 
the corresponding local velocity deflection coefficients generically read
\begin{subequations}
\label{eq:dv_loc}
\begin{align}
\langle \Delta v_i \rangle \!&= \! -2A \! \int\! \rd \bv' \, \frac{w_i}{w^3} \Frot (\br,\bv'),\\
\langle \Delta v_i \Delta v_j \rangle \!&= \! A \! \int\! \rd \bv' \, \frac{w^2 \delta_{ij}-w_i w_j}{w^3} \Frot (\br,\bv'),
\end{align}
\end{subequations}
where ${ 1 \!\leq\! i , j \!\leq\! 3 }$ run over the three directions of the coordinate system. We also  defined ${A \!= \!4 \pi m G^{2} \!\ln \Lambda}$ and the relative velocity ${\bw \!=\! \bv - \bv'}$.
Here, $ \ln \Lambda$ stands for the Coulomb logarithm,
set to ${\Lambda\!=\! 0.11 N}$ as is usual for single-mass globular clusters \citep{Giersz1994,HeggieHut2003}.
In principle, the rewriting of these expressions using Rosenbluth potentials is feasible, as outlined by~\cite{Rosenbluth1957}.
Such a program, in a non-rotating but anisotropic cluster,
was pursued in~\citetalias{Tep2022}.

In appendix~\ref{app:NR}, we improve upon this approach
in the context of rotating clusters.
In particular, we show how a rewriting rather based on equations~\eqref{eq:dv_loc}
improves the numerical stability and ensures the positivity
of the second-order diffusion coefficients.
More precisely, we write equations~\eqref{eq:dv_loc} as
\begin{equation}
\label{eq:dv_loc_ref}
\begin{bmatrix}
\dvPar
\\
\dvParSq
\\
 \dvPerpSq 
 \end{bmatrix}
\!=\! A \!\! \int \!\! \rd w \rd \vartheta \rd \phi \sin \vartheta
\begin{bmatrix}
- 2 \cos  \vartheta
\\
w \sin^2 \vartheta  
\\
w \, (1+\cos^2 \vartheta)    
\end{bmatrix} \Frot ,
\end{equation}
where $ \dvParLoc $ (resp.\ $ \dvPerpLoc$) is the local velocity deflection along (resp. perpendicular to) the test star's trajectory and we have used polar coordinates $(w,\vartheta,\phi)$ with the $z$-axis parallel to $\bv$ (see Fig.~\ref{fig:Coord_syst_1_spe}).
Appendix~\ref{app:LocalDiff} also details the arguments at which the \DF, ${\Frot(\br,\bv')\!=\!\Frot(E',L',\Lz')}$, must be evaluated.

Equation~\eqref{eq:dv_loc_ref} is one of the main results
of the present work.
In particular, equation~\eqref{eq:dv_loc_ref}
possesses a few key advantages compared to the expressions
given in equations~{(3)} of \citetalias{Tep2022},
which are recovered by the present approach:
(i) no integrable singular denominators remain;
(ii) no gradients of $\Frot$ are required during the computation;
(iii) the positivity of the second-order coefficients is ensured;
(iv) this equation applies to a wider range of clusters. 
Ultimately, equation~\eqref{eq:dv_loc_ref} compactly accounts
for all the two-body deflections from the cluster's stars onto the test star.

\subsection{Orbit average and secular evolution}
\label{sec:OA_SE}

Because we are interested in the diffusion of orbits,
we must consider the effect of \NR\ on orbital invariants,
namely the energy $E$ and the angular momenta $L$ and $L_{z}$.
In appendix~\ref{sec:Lz_coeffs}, we expand upon~\citetalias{Tep2022}
and detail how the local diffusion coefficients in ${ (E,L,\Lz) }$
may be computed from the local velocity coefficients given in equation~\eqref{eq:dv_loc_ref}.

\begin{figure} 
    \centering
   \includegraphics[width=0.3\textwidth]{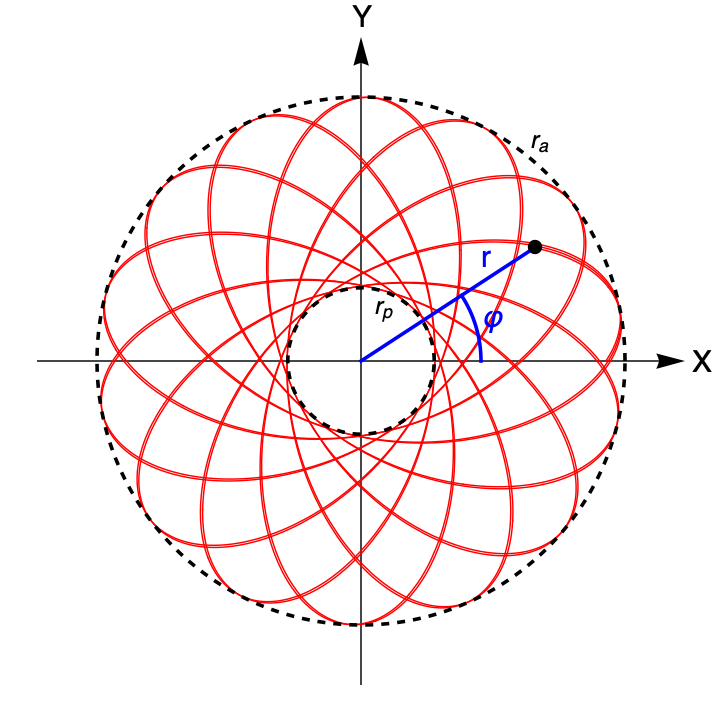}
   \caption{Illustration of the orbit average in physical space.
   The test star (in black, part of whose orbit is shown in red) is averaged over all its available positions in its orbital plane,
   within the radial range of the orbit (${r\! \in\! [\rrp,\rra]}$) and its angular range ($\varphi \!\in\! [0,2\pi]$),
   weighted by the surface density $\Omega_r/(2\pi^2 |\vr|)$.
   Here, the frame is that of the right panel of Fig.~\ref{fig:coordinate_system}.
      }
   \label{fig:orbit_average}
 \end{figure}

Once the local diffusion coefficients are known,
they can be averaged along the unperturbed mean field orbit of the test star.
At this stage, our accounting of rotation adds some complexity.
Indeed, the orbit average now involves an intricate two-dimensional integral
spanning the radial range of the test orbit (${r\! \in\! [\rrp,\rra]}$)
and its angular phase ($\varphi \!\in\! [0,2\pi]$) within the orbital plane.
This is illustrated in Fig.~\ref{fig:orbit_average}.

In practice, computing the orbit average diffusion coefficients
involves computing expressions of the form
\begin{equation}
\label{eq:generic_orbit_average}
D_{X} = \frac{\Omega_r}{\pi} \int_{\rrp}^{\rra} \frac{\rd r}{|\vr|} \int_{0}^{2\pi} \frac{\rd \varphi}{2\pi} \, \langle \Delta X \rangle (r,\varphi),
\end{equation}
where $X$ runs over all the diffusion coefficients to compute,
that is, runs over singlets and pairs from ${ \{ E , L , \Lz \} }$.
Here, $\Omega_r$ is the radial frequency and $\vr$ the radial velocity
that is evaluated along the mean field orbit.
We refer to appendix~\ref{app:orbit_average}
for detailed expressions.
In particular, in appendix~\ref{app:orbit_average_angle},
we show how the angular average
can be performed explicitly for some classes of rotating models.
And, in appendix~\ref{app:orbit_average_radial},
we use the same approach as in appendix~{F2} of~\citetalias{Tep2022}
to compute the radial average using
a stable numerical integration scheme.

The final stage of the calculation is to convert
the orbit-averaged diffusion coefficients in $(E,L,\Lz)$
into ones in ${\bJ\!=\!(\Jr,L,\Lz)}$.
This is rather straightforward,
as detailed in appendix~\ref{app:action_coeffs}.
Lastly, these diffusion coefficients are those
used to evaluate the \FP\ equation~\eqref{eq:def_FP}.
Along the same lines, one can similarly describe
the dynamics within the coordinates ${\bJc\!=\!(\Jr,L,\cos I)}$.

\section{Long-term relaxation}
\label{sec:LongTerm}

We wish to study the impact of rotation on the long-term evolution
of rotating, spherically symmetric\footnote{We will justify the assumption of spherical symmetry in section~\ref{sec:sphericality}.}, anisotropic clusters.
In this first section, we focus on direct $N$-body simulations
to get some insight on the dynamics at play.
We detail the corresponding numerical setup in appendix~\ref{sec:NBODY}.

First, let us describe the classes of clusters considered.
The in-plane distribution of orbits (i.e.,\ when projecting in the ${ (\Jr,L) }$ space)
follows the same Plummer \DFs\ \citep[see, e.g.,][]{Dejonghe1987} as in section~{3.1} of~\citetalias{Tep2022}.
Velocity anisotropy is encoded
with the dimensionless parameter $q$,
with $q=0$ corresponding to an isotropic distribution,
and ${ q \!>\! 0 }$ (resp.\ ${ q \!<\! 0 }$)
associated with radially (resp.\ tangentially) anisotropic velocity distributions.

Let us introduce rotation in these models,
while leaving the mean potential invariant.
To do so, we follow the \LBD\ \citep{LyndenBell1960}
and consider
\begin{equation}
\label{eq:LBD}
\Frot(\bJ) = \Ftot(\Jr,L) \, \big(1+\alpha \,\sgn[\Lz/L] \big),
\end{equation}
where $\Ftot(\Jr,L)$ is the non-rotating \DF\
and $\alpha$ a dimensionless parameter between 0 and 1.
Physically, this parameter corresponds to converting a fraction $\alpha$
of retrograde orbits ${(\Lz\!<\!0)}$
into prograde orbits ${(\Lz\!>\!0)}$.
Importantly, since ${ \Lz \!\mapsto\! \sgn[\Lz/L] }$
is an odd function,
we stress that $\Ftot$ and $\Frot$
generate the exact same potential.

When considering orbital inclinations
via ${ \bJc \!=\! (\Jr , L , \cos I) }$
(see appendix~\ref{app:convert_Lz_to_cosI}),
the \LBD\ yields the DF in $\bJc$ space  \footnote{Using $\cos I$ as an effective coordinate
to describe relaxation might be problematic for some clusters.
We refer to Appendix~\ref{app:impact_disct} for more details.}
\begin{equation}
F (\bJc) =  L \, \Ftot (\Jr , L) \, \big( 1 \!+\! \alpha \, \sgn [\cos I] \big) .
\label{eq:LBD_cosI}
\end{equation}
Integrating this equation over $\cos I$,
one recovers the reduced \DF\ in ${ (\Jr,L) }$,
${F(\Jr,L)\! =\! 2 L \Ftot(\Jr,L)}$,
used to describe non-rotating clusters~\citep[see, e.g.,][]{Hamilton2018}.

\subsection{Sphericity of the cluster}
\label{sec:sphericality}

The \NR\ theory presented in section~\ref{sec:NR}
assumes that the cluster remains spherically symmetric throughout.
However, \cite{Rozier2019} showed that (sufficiently) rotating clusters
can harbour unstable modes (see, e.g., fig.~{8} therein).
To avoid this complication, we  restricted ourselves
to (linearly) stable rotating clusters,
that is, clusters that would remain spherically symmetric.

\begin{figure} 
    \centering
   \includegraphics[width=0.45 \textwidth]{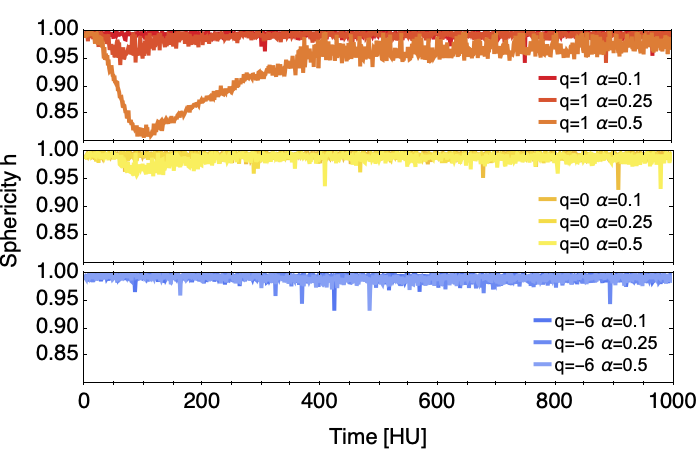}
   \caption{Sphericity, $h$, of a sample of rotating anisotropic clusters,
   as defined in appendix~\ref{app:sphericality},
   from $N$-body simulations for ${N\!=\!10^5}$.
   Each case has been averaged over $50$ realisations. We represent radially anisotropic  clusters ${(q\!=\!1)}$ in red, isotropic clusters  ${(q\!=\!0)}$  in yellow,
    and tangentially anisotropic  ones ${(q\!=\!-6)}$ in blue.
    For each anisotropy, we consider three rotating parameters ${\alpha\!=\!0.1,0.25,0.5}$
    (ordered from dark to light colors).
   In agreement with~\citet{Rozier2019},
   some of these clusters are unstable
   -- namely ${(q,\alpha)\!=\!(1,0.5)}$, and in a smaller fashion  ${(q,\alpha)\!=\!(1,0.25)}$ --
   while the others are linearly stable and remain spherically symmetric. }
   \label{fig:sphericality_rotation}
 \end{figure}

To probe the conservation of spherical symmetry,
we investigateed the ``sphericity'', $h$, of clusters,
as defined in appendix~\ref{app:sphericality} in terms of
the eigenvalue ratio of
the density-weighted inertia tensor.
This is illustrated in Fig.~\ref{fig:sphericality_rotation}.
Except for the radially-anisotropic and rapidly-rotating cluster ${ (q , \alpha) \!=\! (1,0.5) }$,
all clusters remain approximately spherical. This is in agreement with the measurements from \citet{Rozier2019}, as this particular cluster falls into the region of linear instability.
Therefore, we shall focus our interest on parameters $(q,\alpha)$
for which the clusters remain spherically symmetric.

\subsection{The gravo-gyro catastrophe}
\label{sec:gravo_gyro}

The typical relaxation timescale of a globular cluster
can be estimated through the half-mass relaxation time~\citep[see, e.g., section~{14} of][for more details] {HeggieHut2003}, defined by 
\begin{equation}
\label{eq:trh}
\trh = \frac{0.138 \,N\,  \rh^{3/2}}{(G M)^{1/2} \ln (0.11 \,N)},
\end{equation}
with $\rh$ the half-mass radius.
For a cluster with ${N\!=\!10^5}$ stars, this yields ${\trh \!=\! 994 \, \rHU}$, where $\rHU$ stands for the H\'enon units~\citep{HeggieMathieu1986}
in which ${G \!=\! M \!=\! \Rv \!=\! 1}$, with $\Rv$ the virial radius\footnote{For a Plummer cluster,
this is related to the Plummer scale length $b$ by the relation ${\Rv\!=\!16b/(3\pi)}$
\citep[see, e.g.,][]{Dejonghe1987}.}.
In practice, with ${ N \!=\! 10^{5} }$ running
a numerical simulation up to ${ t \!\sim\! \trh }$
took about one day of computation, see appendix~\ref{sec:NBODY}.
In \cite{Breen2017}, core collapse occurs at ${ t \!\sim\! \!17 \trh }$ for an isotropic cluster
(see table~{1} therein for the dependence with respect to velocity anisotropy).
Hence, integrating clusters with ${ N \!=\! 10^{5} }$
up to core collapse is not reasonably feasible.
In this section we therefore scaled down
the size of the clusters to ${N\!=\!10^4}$ stars.
In that case,  ${\trh \!=\! 132 \, \rHU}$,
and core collapse was numerically reached
in about ten hours.
Since we focused on measuring the core radius
-- a very integrated quantity --
the quality of the numerical measurements
was not too much degraded by this use
of a smaller value for $N$.
In practice, measurements were averaged over 50 runs.

\begin{figure} 
    \centering
   \includegraphics[width=0.45 \textwidth]{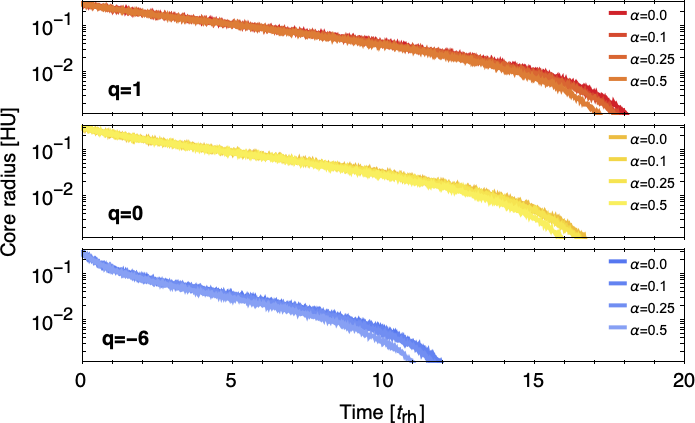}
   \caption{Core radius as a function of time,
   as measured in $N$-body simulations, with $\trh$ defined in equation~\eqref{eq:trh}.
   Each case has been averaged over 50 realisations with ${ N\!=\!10^4 }$.
   Increasing the rotation strength $\alpha$ slightly reduces the time of core collapse.
   Nevertheless, the impact of rotation (i.e.\ the gravo-gyro catastrophe)
   is surely not as pronounced as what was observed in, e.g.,~\citet{Einsel1999}. 
   }
   \label{fig:Rc_rotation_N1e4}
 \end{figure}

Figure~\ref{fig:Rc_rotation_N1e4} represents the time evolution
of the averaged core radius, $R_{\rc}$, defined as (see, e.g., \citealt{Breen2017}; \citetalias{Tep2022})  \begin{equation}
R_{\rc}^2 = \sum_i r_i^2 \rho_i^2 / \sum_i \rho_i^2 ,
\end{equation}
where $r_i$ is the radial position of star $i$ and $\rho_i$ an estimator of the density at $r_i$,
as defined in {\sc \small NBODY6++GPU}~\citep{Wang2015}.
for the usual sets of anisotropies and rotations (${q\!=\!1,0,-6}$ and ${\alpha\!=\!0,0.1,0.25,0.5}$).
Interestingly, for these rotating clusters,
the impact of rotation is definitely not as important
as the one reported in~\citet{Einsel1999} in rotating King models
(see figure~2 therein).
Appendix~\ref{app:King} reproduces the main parameters
characterising these models.
For example, in~\citet{Einsel1999},
assuming ${ W_0 \!=\! 6.0 }$, a non-rotating King cluster collapses
at ${t \sim 12 \trh}$,
while a rotating one, with ${\omega_0 / \Omega_{0} =0.4}$,
can collapse as early as ${t \sim 9 \trh}$.
Here, for Plummer spheres,
we do not find any such stark impact of rotation.

Nonetheless, Figure~\ref{fig:Rc_rotation_N1e4} does not contradict
the measurements from~\cite{Einsel1999}.
Indeed, the parameters $\omega_0$ (King) and $\alpha$ (Plummer)
do not parametrise rotation in the same way.
As \cite{Einsel1999} varies $\omega_0$,
the mean density profile gets modified.
Indeed, the rotating \DF\ of the King model (equation~\ref{eq:def_DF_King})
cannot be decomposed into a fixed, rotation-free even part,
and an odd part in $\Lz$ \citep[see, e.g.,][]{Dejonghe1986}.
As such, we argue that the \LBD\  approach
ensures a fairer comparison between models
to isolate the distinctive impact of rotation.

Of course, to better assess the universality of these observations,
it would be worthwhile to perform similar experiments
with other parametrisations for the rotation
(see Appendix~\ref{app:impact_disct}).
This is left for future investigations.

\subsection{In-plane vs out-of-plane diffusion}
\label{sec:LongTerm_inclination}

Using once again $N$-body simulations with ${ N \!=\! 10^{4} }$ stars,
let us finally investigate the typical features
of the in-plane diffusion 
(i.e.\ relaxation in ${ (\Jr, L) }$)
and out-of-plane diffusion 
(i.e.\ relaxation in ${ \cos I}$).

\begin{figure*} 
  \centering
       \includegraphics[width= \textwidth]{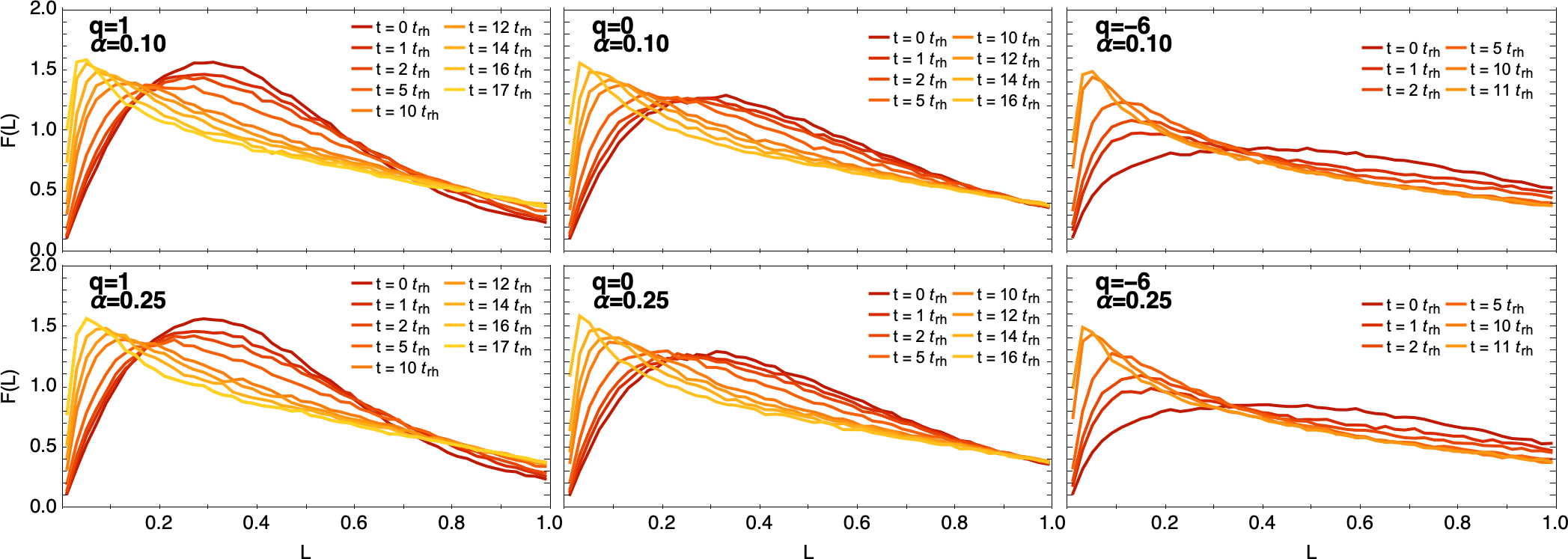} 
   \caption{Evolution of the angular momentum \DF, $F(L)$, as measured in $N$-body simulations with ${N\!=\!10^4}$, with $\trh$ defined in equation~\eqref{eq:trh}.
      For each parameter, the latest time is close to the time of core collapse.
   Each panel is averaged over 50 realisations.
   Here, ${ F (L) }$ redistributes towards lower angular momenta,
   with details depending on the initial velocity anisotropy.
  Rotation does not impact strongly the in-plane relaxation.
   }
   \label{fig:DF_NBODY_L}
 \end{figure*}

First,  Fig.~\ref{fig:DF_NBODY_L}
 reports on the time evolution of the \DF\ in $L$,
that is, the in-plane relaxation of the distribution of angular momenta.
There, we recover the imprints of core collapse visible
through the slow overall contraction of the distribution.
But importantly, we clearly note that rotation only (very) weakly impacts
this relaxation.

\begin{figure*} 
  \centering
   \includegraphics[width= \textwidth]{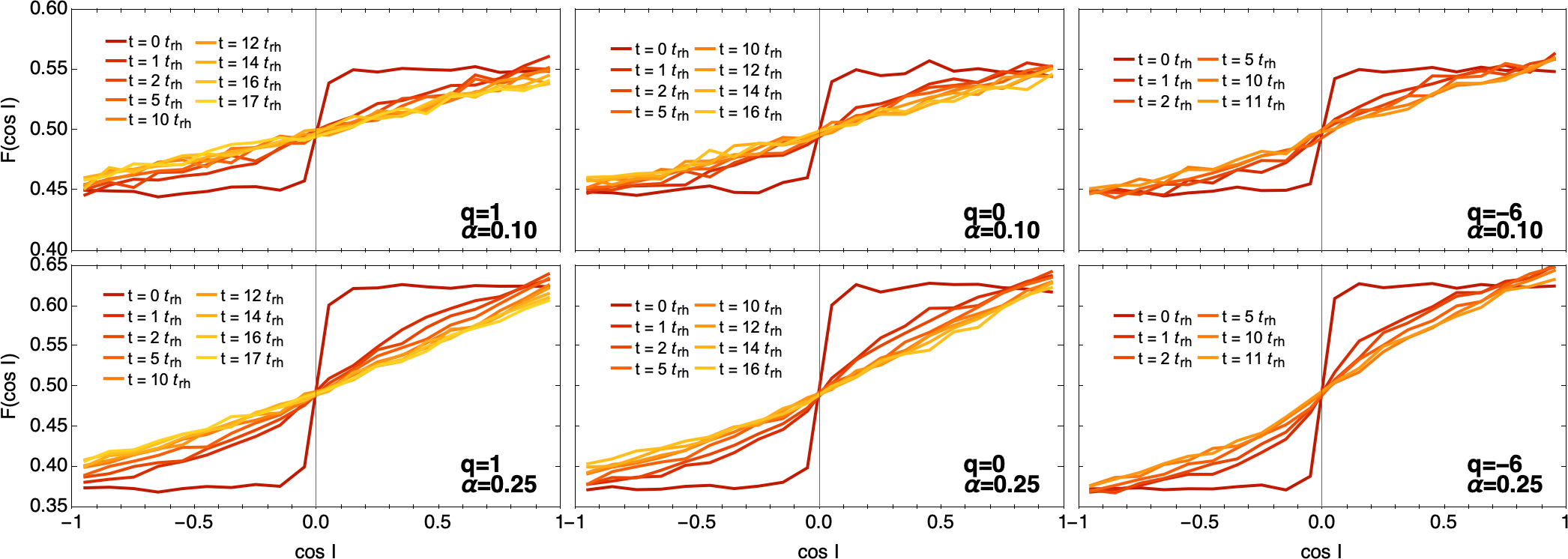}
   \caption{Same as Fig.~\ref{fig:DF_NBODY_L}
   but for the \DF\ in orbital inclinations, ${ F (\cos I) }$.
   The out-of-plane relaxation does not depend much on the initial velocity anisotropy.
   As time evolves, the discontinuity of the \LBD\ (equation~\ref{eq:LBD})
   at ${\cos I \!=\!0}$ is (rapidly) washed out.
      }
   \label{fig:DF_NBODY_CosI}
 \end{figure*}

We can now compare this relaxation with the out-of-plane one.
Figure~\ref{fig:DF_NBODY_CosI}
reports on the time evolution of the \DF\ in $\cos I$.
As could have been expected,  the clusters
tend towards smoother distribution of inclinations,
meaning that relaxation tends to erase discontinuities.
The sharp discontinuity at ${\cos I\!=\!0}$ has already been washed away after $1 \,\trh$, regardless of the initial amplitude of the discontinuity. We refer to appendix~\ref{app:lbd_v_erf} for a comparison of the early relaxation between a discontinuous and a smooth
distribution of orbital inclinations.
Note that
the redistribution of prograde and retrograde orbits
observed in Fig.~\ref{fig:DF_NBODY_CosI} does not conflict
with the conservation of the cluster's total angular momentum.
Indeed, the average of ${\cos I\!=\!\Lz/L}$ is not constrained by any global invariance,
contrary to $\Lz$.
As such, even if the number of particles with positive and negative $\cos I$
(hence $\Lz$) changes,
the conservation of the total angular momentum is ensured
by a modulation of the norm of each star's angular momentum vector.

Comparing Figs.~\ref{fig:DF_NBODY_L} and~\ref{fig:DF_NBODY_CosI}
shows that  orbital inclinations relax much faster
than angular momenta.
This is a point already raised in 
  \citet{Rauch1996} (fig.~{2} therein).
In particular, they showed in their section~{1.4}
that the long-term relaxation of $E$ and $L$ in spherical potential was driven by the \NR\ theory, whereas that of the angular momentum vector $\bL$ -- and thus $\Lz$ -- was subject to an enhanced relaxation.
This out-of-plane relaxation is driven by coherent torques between orbits
and is coined \VRR\@.

\section{In-plane diffusion}
\label{sec:Secular_Jr}

Assuming that the cluster's relaxation is driven by local pairwise deflections,
its long-term evolution is described by the \FP\ equation~\eqref{eq:def_FP}. While this equation formally describes the evolution in 3D action space, it is convenient to study relaxation  in two-dimensional projections,
in particular to improve the signal-to-noise ratio.
Yet, such a projection comes at the cost
of more intricate theoretical predictions,
that require additional integrations along some third action.
In this section, following the same approach as in~\citetalias{Tep2022},
we first focus on in-plane relaxation, that is, relaxation occurring in ${ (\Jr,L)}$.
Our goal is to compare quantitatively the $N$-body measurements
with the \NR\ prediction.

\subsection{$N$-body measurements}
\label{sec:Nbody_measures}

\begin{figure} 
    \centering
   \includegraphics[width=0.45 \textwidth]{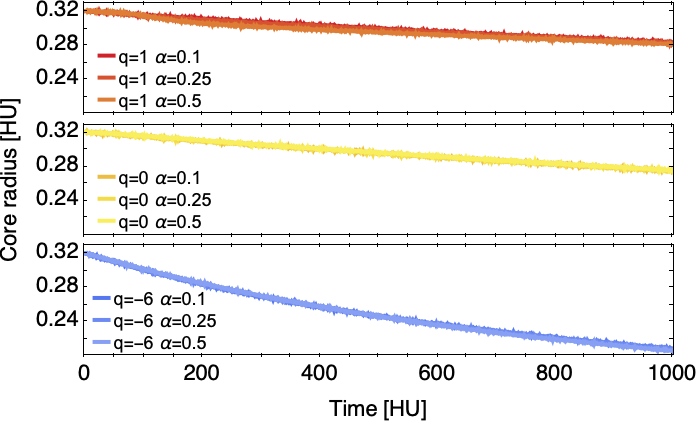}
   \caption{Initial evolution of the core radius in clusters
   with ${ N \!=\! 10^{5} }$ stars, as measured in $N$-body simulations.
   We use that same convention as in Fig.~\ref{fig:sphericality_rotation}.
   Interestingly, in these stable clusters,
   rotation only weakly affects the core contraction.
   }
   \label{fig:Rc_rotation}
 \end{figure}

The onset of core collapse is illustrated in Fig.~\ref{fig:Rc_rotation}
for clusters with ${N\!=\!10^5}$ stars. 
As in Fig.~\ref{fig:Rc_rotation_N1e4},
we recover that rotation
does not really impact the rate of core collapse
in these linearly stable clusters.
Conversely, when the clusters are unstable,
such as for ${ (q , \alpha) \!=\! (1 , 0.5) }$ (see Fig.~\ref{fig:sphericality_rotation}),
we find numerically (not reported in Fig.~\ref{fig:Rc_rotation}) that the clusters flatten
and that relaxation is accelerated.

To probe more precisely relaxation,
let us now measure the relaxation rate, ${ \p F / \p t }$, in ${ (\Jr , L) }$.
This is presented in Fig.~\ref{fig:dFdt_NBODY_JrL}
with details spelled out in Appendix~\ref{sec:NBODY}.
\begin{figure*} 
    \centering
    \includegraphics[width=0.80 \textwidth]{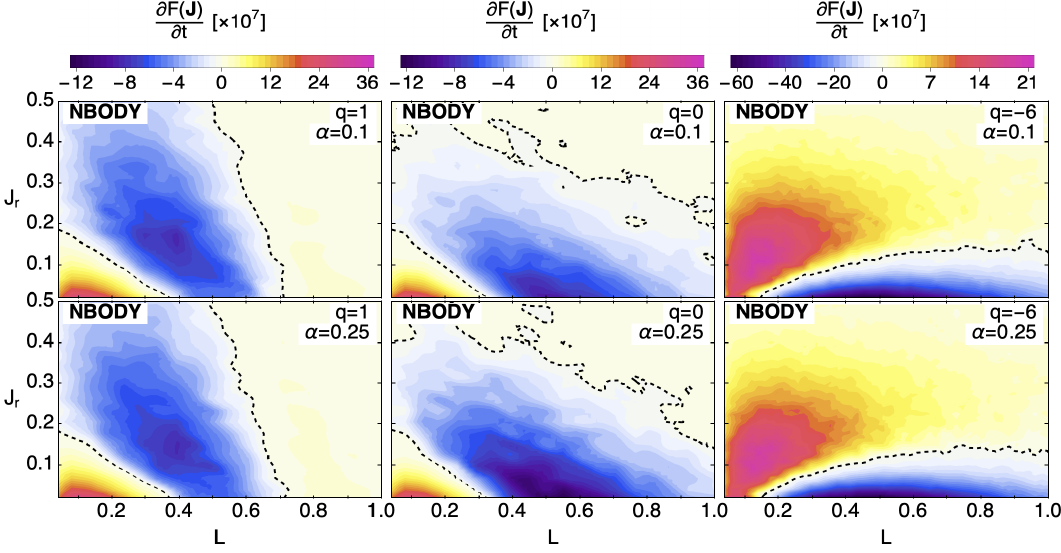}
   \caption{Relaxation rate, $\p F/\p t$, in $(\Jr,L)$
   for various velocity anisotropies $q$ (left to right) and rotation parameters $\alpha$ (top and bottom),
   as measured in $N$-body simulations.
The overall amplitude of the in-plane relaxation rate
   depends on $q$,
   but only very weakly on $\alpha$.  
   }
   \label{fig:dFdt_NBODY_JrL}
 \end{figure*}
Even in the presence of rotation,
we recover the same result as in~\citetalias{Tep2022},
namely all clusters seem to isotropise toward an in-plane \DF\,
depending only on energy. Indeed, in radially anisotropic clusters,
relaxation depletes radial orbits.
And the converse holds for tangentially anisotropic clusters.
Importantly, the amount of rotation, $\alpha$,
has no significant impact on the geometry of the in-plane diffusion.
Increasing $\alpha$ only (very) weakly accelerates the in-plane relaxation.
This is in concordance with the slightly shorter core collapse time
observed in Fig.~\ref{fig:Rc_rotation_N1e4}.

\subsection{Non-resonant prediction}
\label{sec:NR_pred}

Let us now compute the \NR\ prediction
for the relaxation rate in ${ (\Jr,L)}$.
This prediction was numerically very intensive
and required the evaluation of five embedded integrals,
as demonstrated in Appendices~\ref{app:NR} and~\ref{app:2D_FP}.
Importantly, we stress that in the present case,
the \LBD\ (equation~\ref{eq:LBD})
allowed us to analytically perform the angular part of the orbit average (equation~\ref{eq:generic_orbit_average}).
This helped reducing the computation time.
We refer to Appendix~\ref{app:orbit_average_angle} for details.

\begin{figure*}
   \centering
         \includegraphics[width=0.80 \textwidth]{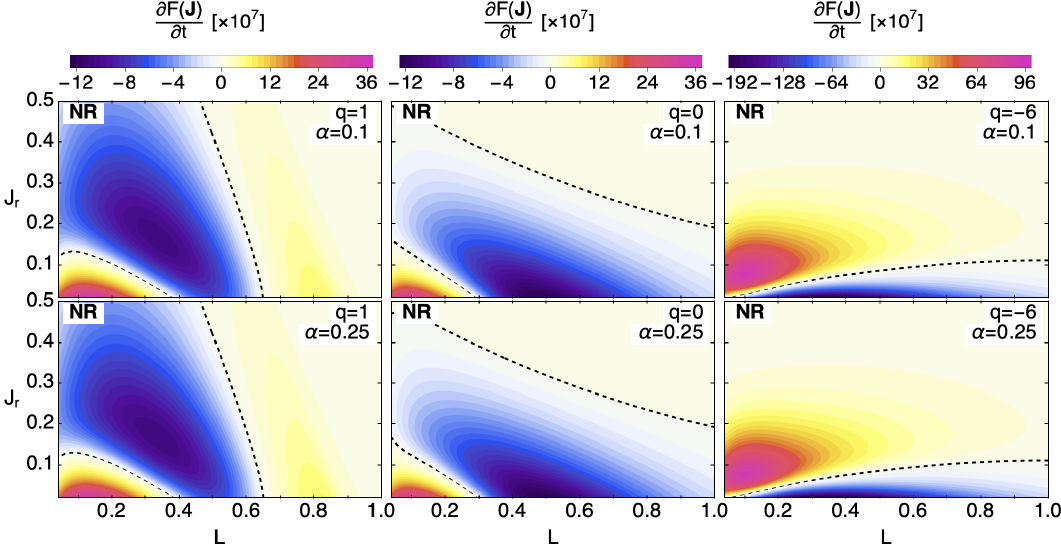}
   \caption{Relaxation rate, $\p F/\p t$, in ${ (\Jr , L) }$
   as in Fig.~\ref{fig:dFdt_NBODY_JrL},
   but here from the \NR\ theory.
   The prediction matches the $N$-body measurements from Fig.~\ref{fig:dFdt_NBODY_JrL},
   up to an overall prefactor that depends on $q$, and (very) weakly on $\alpha$.
   }
   \label{fig:dFdt_NR_JrL}
 \end{figure*}

The relaxation rate, ${ \p F / \p t }$ in ${ (\Jr , L) }$,
presented in Fig.~\ref{fig:dFdt_NR_JrL}
for various anisotropies and rotations,
suggests once again an orbital reshuffling toward isotropisation.
In addition, increasing the rotation parameter $\alpha$ only very weakly increases
the relaxation rate in $(\Jr,L)$,
while barely impacting the structures observed in the non-rotating case.

When comparing the $N$-body measurements (Fig.~\ref{fig:dFdt_NBODY_JrL})
with the \NR\ prediction (Fig.~\ref{fig:dFdt_NR_JrL}), the \NR\ theory successfully recovers the in-plane relaxation
of stable rotating clusters:
both figures exhibit strikingly similar structures in action space.
We obtain results for the non-rotating case that are in agreement with \citetalias{Tep2022}.
Yet, even though amplitudes in both approaches are comparable,
there is still a (slight) overall prefactor mismatch.
The inclusion of rotation has little impact on this mismatch.
A more quantitative comparison would require performing (many) more $N$-body runs, 
as well as a more precise measurement of the relaxation rate. 
This would be no light undertaking, and will be the subject of future works.

\section{Out-of-plane diffusion}
\label{sec:Secular_cosI}

Because they rotate, the present clusters
also undergo some out-of-plane diffusion,
namely  relative to the orbital inclinations $\cos I$. 
In this section,
we focus on relaxation in ${ (\Jr , \cos I) }$,
and we refer to appendix~\ref{app:L_cosI_diff}
for a similar investigation in ${ (L , \cos I) }$.

\subsection{$N$-body measurements}
\label{sec:Nbody_cosI}

\begin{figure*} 
  \centering
   \includegraphics[width=0.80 \textwidth]{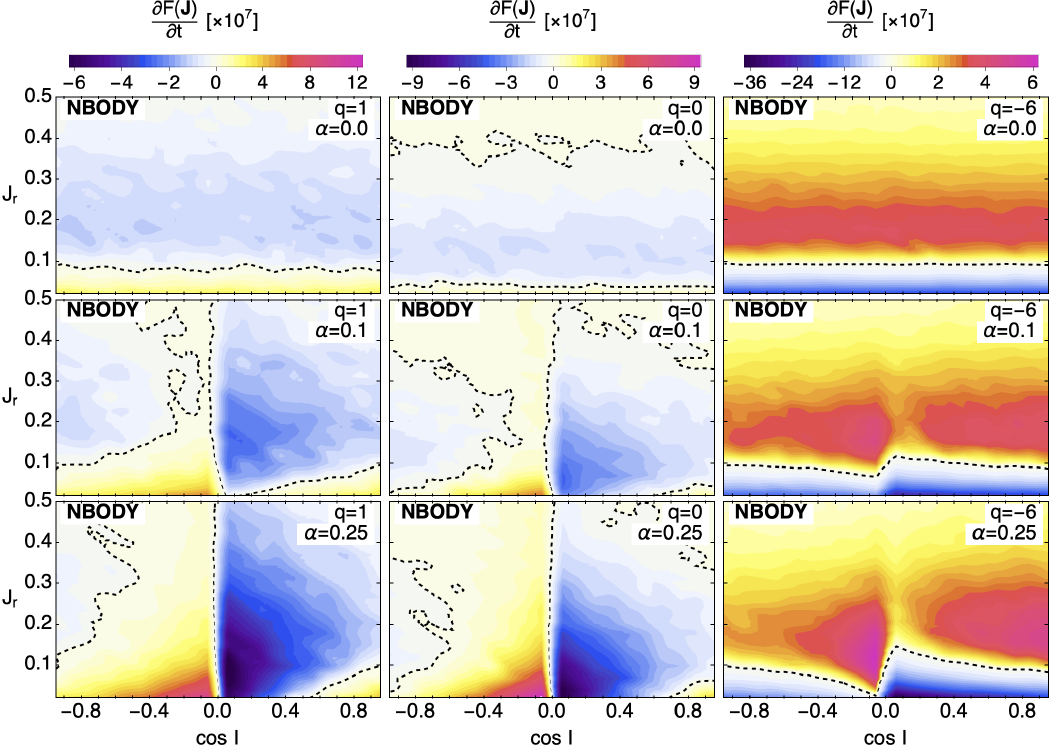}  
   \caption{Relaxation rate, $\p F/\p t$, in ${ (\Jr , \cos I) }$ for various anisotropies $q$ (left to right) and rotation $\alpha$ (from top to bottom), as measured in $N$-body simulations.
   The amplitudes and structures observed depend on anisotropy,
   and show a reshuffling towards isotropisation.
  Orientations redistribute toward a more affine distribution in $\cos I$. As a result of using binning
  in action space and finite difference in time to compute $\p F/\p t$, the expected initially sharp relaxation occurring at $\cos I\!=\!0$ has been smoothed and is not visible. 
   }
   \label{fig:dFdt_NBODY_JrCosI}
 \end{figure*}

The relaxation rates in $N$-body simulations 
are presented in Fig.~\ref{fig:dFdt_NBODY_JrCosI}.
As expected, in the absence of rotation (top row),
the relaxation is independent of $\cos I$,
and the distribution of orientations remains uniform.

Let us now consider the rotating clusters
presented in the middle and bottom rows of Fig.~\ref{fig:dFdt_NBODY_JrCosI}.
In radially anisotropic (${ q \!=\! 1 }$) and isotropic (${ q \!=\! 0 }$) clusters,
the systems lose stars in the prograde region (${\cos I\!>\!0}$)
and gain stars in the retrograde region (${\cos I\!<\!0}$).
As one increases $\alpha$ (i.e.\ as one increases the net rotation),
this trend strengthens.
This is in agreement with the orbital reshuffling observed in Fig.~\ref{fig:dFdt_NR_JrL},
where the clusters always tend to isotropise their in-plane distribution.
Finally, we note that the highest diffusion rates in inclination are observed near ${\cos I\!=\!0}$.
This corresponds precisely to the location
of the discontinuity of the  \LBD\@.

At first glance, the tangentially anisotropic case (${ q \!=\! -6 }$),
as given by the right column in Fig.~\ref{fig:dFdt_NBODY_JrCosI},
may seem different. Indeed, in that case  a depletion
of orbits for small $\Jr$ is  observed, whatever $\cos I$.
Conversely, the number of orbits systematically increases for large $\Jr$,
whatever $\cos I$.
This is directly linked to the in-plane isotropisation of the cluster
(Fig.~\ref{fig:dFdt_NR_JrL}).
Indeed, these clusters being tangentially biased,
the in-plane diffusion occurs towards higher $\Jr$.
Overall, all clusters, independently of their anisotropies,
evolve towards smoother distribution of orbital inclinations.

As a complement of Fig.~\ref{fig:dFdt_NBODY_JrCosI},
appendix~\ref{app:L_cosI_diff} also presents
the relaxation rates in ${ (L , \cos I) }$.
It reaches similar results as for diffusion in ${ (\Jr , \cos I) }$.
Namely, the in-plane distributions tend to isotropise
and orbital inclinations diffuse so as to reach smoother distributions. 

In addition, we stress that due to the use of binning and finite differentiation to compute $\p F/\p t$, the expected sharp evolution occurring at $\cos I \!=\!0$ has been smoothed out. Furthermore, the amplitude of relaxation depends on the time interval used for finite differentiation and on the bin size. While this does not drastically change the observations away from $\cos I \!=\!0$,
any measurement in the neighborhood of $\cos I \!=\!0$ is tricky to perform:
it should be taken with appropriate caution.

\subsection{Non-resonant prediction}
\label{sec:NR_pred_cosI}

\begin{figure*} 
   \centering
     \includegraphics[width=0.80 \textwidth]{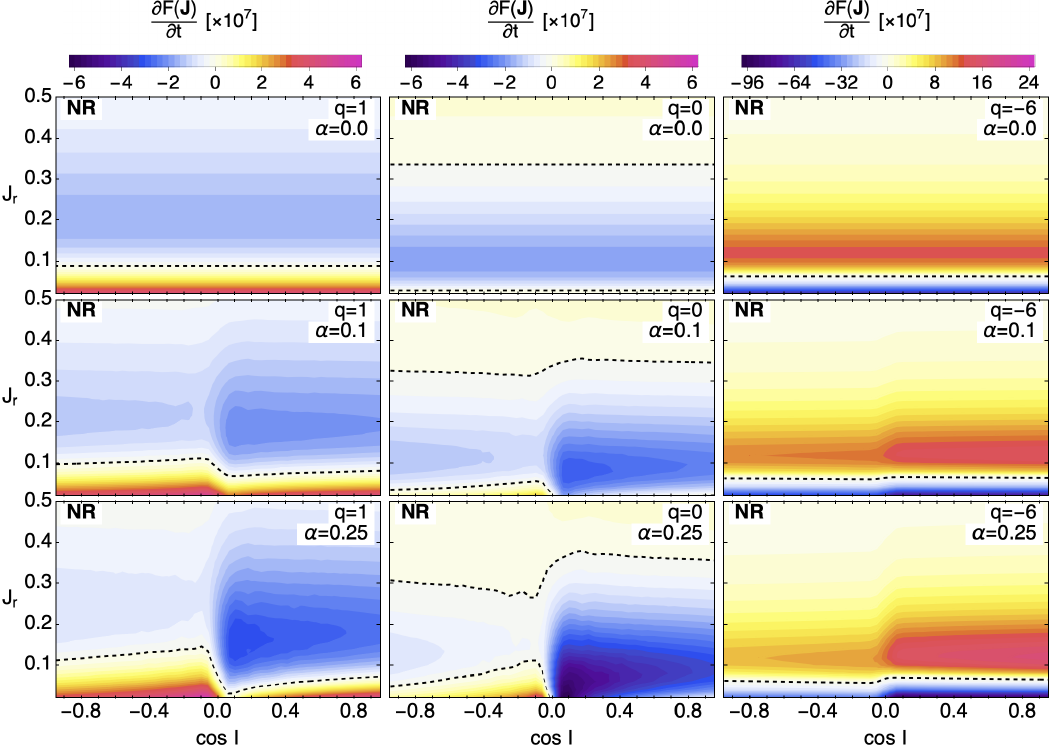}  
   \caption{Relaxation rate, $\p F/\p t$, in ${ (\Jr , \cos I) }$ as in Fig.~\ref{fig:dFdt_NBODY_JrCosI},
   but here from the \NR\ theory (see Appendix~\ref{app:FP_2d_LcosI} for details).
   While the in-plane isotropisation and the smoothing of the inclination distribution
   is recovered, finer details disagree. Here, the amount of relaxation was estimated using finite differentiation using separately regions with positive or negative ${ \cos I }$
   In effect, this removes the $\delta'(\cos I)$ singularity in $\cos I\!=\!0$,
   as discussed in appendix~\ref{app:disc_cosI_0}.
   Importantly, the removal of this singularity does not affect the prediction away from $\cos I=0$. In addition, because this singularity is diluted in $N$-body measurements -- as discussed in Fig.~\ref{fig:dFdt_NBODY_JrCosI} -- we do not represent the \NR\ prediction here.
    }
   \label{fig:dFdt_NR_JrCosI}
 \end{figure*}

Let us now perform the same investigation in ${ (\Jr , \cos I) }$
using the \NR\ theory.
Obtaining a satisfying prediction near ${ \cos I \!=\! 0 }$
required finely sampling the \NR\ orbital integrals,
as detailed in Appendix~\ref{app:FP_2d_LcosI}.
The \NR\ prediction is illustrated in Fig.~\ref{fig:dFdt_NR_JrCosI}.

First, in the absence of rotation,
the \NR\ theory predicts
a relaxation rate independent of $\cos I$:
this is a reassuring sanity check.
Yet, when comparing the top rows of Figs.~\ref{fig:dFdt_NBODY_JrCosI} and~\ref{fig:dFdt_NR_JrCosI},
we note some discrepancies.
Namely, the locations of the line of zero relaxation
(dashed lines) slightly differ.

The difference is much more striking in the presence of rotation
(middle and bottom rows of Fig.~\ref{fig:dFdt_NR_JrCosI}).
In both figures, the orbital reshuffling still operates,
and the clusters redistribute their orbits towards smoother distributions
of inclinations.
Yet, the diffusion structures predicted by the \NR\ theory
do not align with the $N$-body ones.
Indeed, $N$-body simulations (Fig.~\ref{fig:dFdt_NBODY_JrCosI})
present ``round'' structures
as one goes from ${ \cos I \!=\! 0 }$ to ${ \cos I \!=\! 1 }$.
In addition, the relaxation rate decreases as one considers
orbits within to rotation plane,
that is, as one considers ${ \cos I \!\to\! \pm 1 }$.
This is in sharp contrast with the ``straighter'' structures
predicted by the \NR\ theory (Fig.~\ref{fig:dFdt_NR_JrCosI}).
Nevertheless, one should keep in mind that Fig.~\ref{fig:dFdt_NR_JrCosI} cuts out the singularity at ${\cos I\!=\!0}$. The comparison to the $N$-body measurements near ${\cos I\!=\!0}$ is made difficult by the fact that diffusion at early time is very fast  in that region, as shown by Fig.~\ref{fig:ldb_erf}. As a consequence, the measured relaxation rate may suffer from truncation errors -- due to, for example, the finite differentiation scheme -- the closer one gets to ${\cos I\!=\!0}$. However, as Fig.~\ref{fig:ldb_erf} shows, the initial discontinuity appears to have little impact on diffusion both far away from ${\cos I\!=\!0}$ and beyond ${t\!=\!0.1\, \trh}$.
 The obtention of a more robust measure would require both a finer time difference and a finer binning of action space.
 These two improvements would require an ensemble-average over much more realisations to reduce dispersion.
A similar conclusion is reached in appendix~\ref{app:L_cosI_diff}
when considering relaxation in ${ (L , \cos I)}$.

Before concluding, let us elaborate further
on the failure of the \NR\ theory to model out-of-plane relaxation. For ${\alpha \!\neq\!0}$, 
recall that  ${\cos I\!=\!\pm 1}$ corresponds to orbits within the rotation plane,
while ${\cos I\!=\!0}$ corresponds to orbits perpendicular to the rotation plane. A test star orbiting in this rotating cluster will torque with all the other stars of the cluster.
Equivalently, this means that this star will typically be subject to a torque with the total angular momentum of the rotating cluster, ${\langle \bL \rangle \!\propto\! \alpha\, \mathbf{e}_{z}}$, the strength of which  depends on the test star's orbital orientation. This is what we observe in Fig.~\ref{fig:dFdt_NBODY_JrCosI}. 
However,
the \NR\ theory only takes into account local deflections, 
and therefore cannot take these coherent interactions into account.

Following the work of~\citet{Meiron2019} (see also references therein),
we anticipate for \VRR\ to play a key role in these coherent interactions -- even after factoring the truncation errors due to the finite difference scheme and the singularity at ${\cos I\!=\!0}$,
both discussed earlier in this section.
Indeed, \VRR\@, which is driven by persistent torques between orbital planes,
might be needed to explain
(i) the faster out-of-plane   relaxation (section~\ref{sec:LongTerm_inclination});
(ii) the discrepancies between the $N$-body measurements
and the \NR\ predictions.
Investigating quantitatively the efficiency of \VRR\
in these systems will be the topic of future investigations.

\section{Conclusions and perspectives}
\label{sec:discussion}

\subsection{Conclusion}
\label{sec:conclusion}

In this paper, we used the \NR\ formalism
to probe the long-term time-evolution of rotating anisotropic globular clusters.
First, using $N$-body simulations,
we showed how rotation only has a weak impact
on the time of core collapse.
As such, in contrast with prior research~\citep[see, e.g.,][]{Hachisu1979},
we did not observe any gravo-gyro catastrophe
that could expedite core collapse~\citep[see, e.g.,][]{Einsel1999}.
Since we introduced rotation in a different fashion,
this does not contradict previous results.

Focusing on in-plane diffusion,
we showed how the \NR\ prediction successfully recovers
all the intricacies of the $N$-body measurements.
This is in line with the non-rotating results from~\citetalias{Tep2022}.
Yet, although the diffusion structures in action space closely align,
there is still an overall amplitude mismatch
between the $N$-body measurements and the kinetic prediction.

We subsequently turned our interest on out-of-plane relaxation,
that is, the redistribution of orbital inclinations.
We pointed out the similarities and differences
between the $N$-body measurements and the \NR\ prediction.
In both approaches, we observe
a systematic reshuffling of orbital inclinations
from overpopulated regions to underpopulated ones.
As such, the distribution of inclinations get smoother through time,
and this process is accelerated by rotation. On the one hand, we observed that the initial discontinuity at ${\cos I\!=\!0}$ induced a fast relaxation at small times.
Surely, this strongly affects any measurement of the relaxation rate in that region.
On the other hand, for orbits aligned with the cluster's rotation plane -- where diffusion is much slower and measurements are much easier --
the local \NR\ theory does not match with the $N$-body measurements. We argue that this mismatch originates from the fact that the \NR\ formalism is sourced by local deflections,
which neglect the coherent torques between the orbital planes.
Accounting for the (efficient) contribution of persistent torques
requires the use of \VRR\@.

\subsection{Perspectives}
\label{sec:perspectives}

Here, the \NR\ formalism was used to describe the very onset of relaxation.
Yet, the short duration of our simulations and our limited sample of initial conditions
 prevented us from fully assessing the asymptotic late time orbital distribution.
Naturally, it would be interesting to push the $N$-body integrations further in time.
While some studies carried out such long-term $N$-body simulations~\citep{Tiongco2020,Livernois2022},
orbital inclinations were, unfortunately, not their focus.
Along the same line, it would also be of interest to integrate the 3D \FP\ equation itself.
In isotropic clusters, this process has been conducted,
as demonstrated for instance in the work of~\cite{Vasiliev2015}.
However, the scenario involving anisotropic, rotating systems
has yet to be comprehensively investigated.

Ultimately, if rotation is sufficiently large,
it may induce a flattening of the cluster.
In that case, St\"{a}ckel systems~\citep{Dejonghe+1988}
could be used to model realistic flattened rotating structures.
In particular, this would still ensure the integrability of the mean potential,
that is the existence of explicit angle-action coordinates.
Tailoring secular theory to such setups will be the topic of future works.

In section~\ref{sec:Secular_cosI}, we pointed out
how the \NR\ theory does not reproduce the out-of-plane diffusion 
observed in $N$-body simulations.
To understand this mismatch, it would be valuable to examine the predictions of the \RR\ formalism~\citep[see, e.g.\@,][]{Hamilton2018,Fouvry2021}
in the presence of rotation.
While the inhomogeneous Landau equation was already investigated
in non-rotating isotropic clusters~\citep{Hamilton2018,Fouvry2021},
its implementation in rotating spheres with an explicit dependence on $\Lz$
should be the next  step.
Furthermore, to account for collective effects,
an explicit implementation of the Balescu--Lenard equation \citep{Heyvaerts2010}
might prove necessary to match the details of stacked $N$-body measurements.

Similarly, to get a better handle on the mismatch
reported in section~\ref{sec:Secular_cosI},
one should investigate quantitatively the efficiency of \VRR\
in these clusters ~\citep[][]{Rauch1996}. 
This should allow us to determine if indeed coherent
and long-lasting torques between orbital planes
are the missing ingredients to describe the relaxation of orientations.
While this has been extensively studied in galactic nuclei \citep[see, e.g.,][]{Eilon2009,Kocsis2011,Kocsis2015,SzolgyenKocsis2018,Fouvry2019,Szolgyen2019,Szolgyen2021,Magnan2022}, 
the implementation of \VRR\ in globular clusters
should be the topic of future work~\citep[see][for a preliminary investigation]{Meiron2019}.

It would be interesting to leverage multiple masses into the theory,
so as to capture (radial or inclination) segregation effects~\citep{Meiron2019}.
\cite{Dekel+2023} argues for instance that the excess of massive galaxies found at very high redshift by
the \textit{James Webb Space Telescope} could originate from early feedback-free star
formation within dense, possibly rotating stellar clusters made of massive stars. 
The fate of such clusters should be captured by the present secular theory.
In particular, it could lead to the formation of intermediate mass \BHs\@~\citep{Greene+2004,Greene+2020}
and possibly seeds for supermassive \BHs\@~\citep{Kormendy+2013}.

\section*{Data availability}

The data underlying this article 
is available through reasonable request to the authors.
The code, written in {\sc \small julia}~\citep{JuliaCite},
computing the \NR\ diffusion coefficients in anisotropic rotating clusters
is available at the URL: \href{https://github.com/KerwannTEP/CARP}{https://github.com/KerwannTEP/CARP}.

\section*{Acknowledgements}

This work is partially supported by the grant \href{https://www.secular-evolution.org}{\emph{SEGAL}} ANR-19-CE31-0017
of the French Agence Nationale de la Recherche,
and by the Idex Sorbonne Universit\'e.
We also thank the KITP for hosting the workshop
\href{https://www.cosmicweb23.org}{`\emph{CosmicWeb23}'} during which this project was advanced.
This work is partially supported  by the National Science Foundation under Grant No. NSF PHY-1748958 and Grant No. AST-2310362.
We are grateful to M.~Roule, M.~Petersen, A.~L.~Varri and the members of the Segal collaboration
for numerous suggestions during the completion of this work.
We thank St\'ephane Rouberol for the smooth running of the
Infinity cluster, where the simulations were performed.

Finally, we thank our referee, Prof. Douglas Heggie, for his remarks, which contributed greatly to the improvement of this paper.

\begin{appendix}

\section{Non-resonant theory}
\label{app:NR}

In this Appendix, we detail the derivation of equation~\eqref{eq:dv_loc_ref}.
This is the backbone of our computation of the \NR\ theory.

\subsection{Local velocity coefficients}
\label{app:LocalDiff}

\begin{figure} 
\centering
\includegraphics[width=0.3 \textwidth]{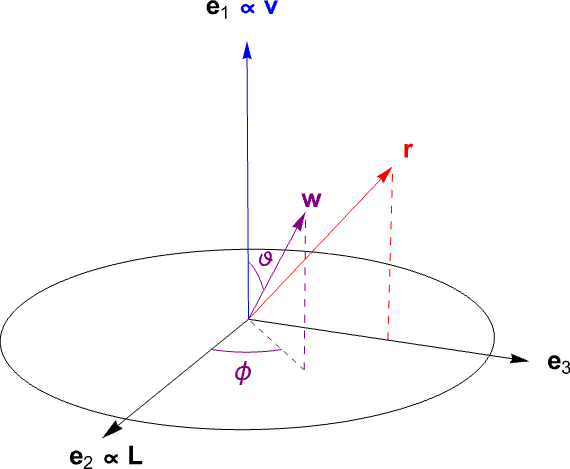}
\caption{Tailored frame used to compute the parallel
and perpendicular local velocity deflections
in equations~\eqref{eq:SpecialCoordSys_LocDiffCoeffs}.
By construction, the test star's angular momentum $\bL$ is along the axis $\mathbf{e}_2$.
}
\label{fig:Coord_syst_1_spe}
\end{figure}

The starting point are equations~\eqref{eq:dv_loc}.
We consider this relation within the frame from Fig.~\ref{fig:Coord_syst_1_spe}.
The local velocity deflections are then given by
\begin{subequations}
\label{eq:SpecialCoordSys_LocDiffCoeffs}
\begin{align}
  \dvPar&=\langle\Delta v_{1}\rangle ,
  \\
  \dvParSq&=\langle(\Delta v_{1})^{2}\rangle ,
  \\
  \dvPerpSq&=\langle(\Delta v_{2})^{2}\rangle+\langle(\Delta v_{3})^{2}\rangle ,
\end{align}
\end{subequations}
where the coordinate 1 is the $z$ coordinate, parallel to $\bv$.
The coordinates 2 and 3 are defined such that the projection of $\br$ on the $(Oxy)$ plane is along $\mathbf{e}_3$ (Fig.~\ref{fig:Coord_syst_1_spe}).
We now introduce ${ \bw \!= \!\bv \!-\! \bv' }$
and change the integration variables from $ \bv'
$ to $\bw$ in equation~\eqref{eq:dv_loc}. 
We get
\begin{subequations}
\label{eq:dv_i}
\begin{align}
\langle \Delta v_i \rangle &= -2A \int \rd \bw \, \frac{w_i}{w^3} \Frot ,\\
\langle \Delta v_i \Delta v_j \rangle  &= A \int \rd \bw \, \frac{w^2 \delta_{ij}-w_i w_j}{w^3} \Frot ,
\end{align}
\end{subequations}
with the shortened notation ${ \Frot \!=\! \Frot (r , \bv' \!=\! \bv \!-\! \bw) }$.
Using these equations allows us to compute
the needed coefficients from equation~\eqref{eq:SpecialCoordSys_LocDiffCoeffs} through
\begin{equation}
\begin{bmatrix}
\langle \Delta v_{1} \rangle
\\
\langle (\Delta v_{1})^{2} \rangle
\\
\langle (\Delta v_{2})^{2} \rangle
\\
\langle (\Delta v_{3})^{2} \rangle
\end{bmatrix}
= A \!\! \int \!\! \frac{\rd \bw}{w^{3}} \begin{bmatrix}
- 2 w_{1}
\\[0.5ex]
w^{2} - w_{1}^{2}
\\[0.5ex]
w^{2} - w_{2}^{2}
\\[0.5ex]
w^{2} - w_{3}^{2}
\end{bmatrix} \Frot
 .
\label{eq:dvi}
\end{equation}
We now define the  spherical
coordinates associated with this frame (Fig.~\ref{fig:Coord_syst_1_spe})
\begin{align}
\label{eq:sphe_w}
w_1 = w \cos \vartheta \,;\, w_2 = w \sin\vartheta  \cos \phi \,;\, w_3 = w \sin \vartheta  \sin \phi.
\end{align}
Injecting equations~\eqref{eq:sphe_w} into equations~\eqref{eq:dvi}, we obtain
\begin{equation}
\label{eq:res_A1}
\begin{bmatrix}
\langle \Delta v_{1} \rangle
\\
\langle (\Delta v_{1})^{2} \rangle
\\
\langle (\Delta v_{2})^{2} \rangle
\\
\langle (\Delta v_{3})^{2} \rangle
\end{bmatrix}
\!=\! A \!\! \int \!\!  \rd w \rd \vartheta \rd \phi  \sin \vartheta 
\begin{bmatrix}
- 2  \cos \vartheta
\\
w \sin^2 \vartheta
\\
 w   (1 - \sin^2 \vartheta  \cos^2 \phi )  
\\
w    (1 -  \sin^2 \vartheta  \sin^2 \phi )  
\end{bmatrix} \Frot .
\end{equation}
Here, the background distribution, $\Frot$,
is to be evaluated in ${\Frot (\br,\bv')\!=\! \Frot (E',L',\Lz') }$.
We compute these arguments in appendix~\ref{app:Arg_Back}.
Finally injecting equation~\eqref{eq:res_A1}
into equation~\eqref{eq:SpecialCoordSys_LocDiffCoeffs},
we obtain our main result,
namely equation~\eqref{eq:dv_loc_ref},
as given in the main text.

\subsection{Arguments of the background distribution}
\label{app:Arg_Back}

Let us now compute the arguments $E'$, $L'$ and $\Lz'$
for the background distribution function, $\Frot (E', L' , \Lz')$,
in equation~\eqref{eq:dv_loc}.

We start with $E'$. It reads
\begin{align}
E' &\!=\! \psi(r) + \frac{\bv^{\prime 2}}{2} \!=\! \psi(r) + \frac{(\bv-\bw)^2}{2}
\!=\! E + \frac{w^2}{2} - w v \cos \vartheta ,
\label{eq:arg_Ep}
\end{align}
since ${\bv\!=\!v\, \mathbf{e}_1}$
(see Fig.~\ref{fig:Coord_syst_1_spe}).
This form allows us to get an upper bound on the $w$-integral of equation~\eqref{eq:res_A1},
above which ${ E' \!>\! 0 }$.
As soon as ${ E' \!>\! 0 }$,
the background star is unbound
and its \DF\ vanishes.

\begin{figure}
\centering
\includegraphics[width=0.45 \textwidth]{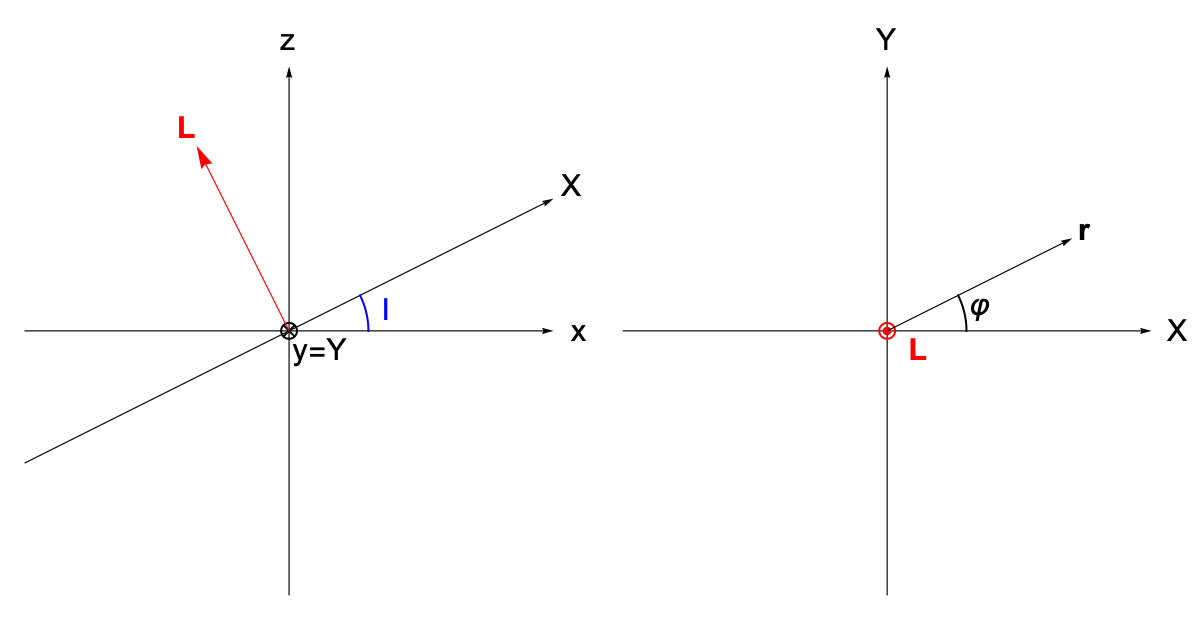}
\caption{Illustration of the frames used to compute the argument of the background DF, $\Frot(E',L',\Lz')$. 
In the left panel, the cluster is seen from the side of the rotation plane, $(Oxy)$. In the right panel, the $(OXY)$-plane is the orbital plane of the test star, seen from above.
}
\label{fig:coordinate_system}
\end{figure}

The computation of $L'$ and $\Lz'$ is best achieved by using a different coordinate system,
as illustrated in Figs.~\ref{fig:Coord_syst_1_spe} and~\ref{fig:coordinate_system}.
More precisely, the frame~$(1,2,3)$ is related to the $(X,Y,Z)$ one through the relations
\begin{equation}
\label{eq:XYZ_to_123}
\left[\hspace{-2mm}\begin{array}{l}
\mathbf{e}_1\\
 \mathbf{e}_2 \\
 \mathbf{e}_3
\end{array}\hspace{-2mm}\right]=  \left[   \begin{array}{ccc}
\frac{\vr}{v} \cos \varphi - \frac{\vt}{v} \sin \varphi   &\frac{\vr}{v}\sin\varphi  + \frac{\vt}{v} \cos \varphi& 0 \\
0&0&1 \\
\frac{\vr}{v}\sin \varphi + \frac{\vt}{v} \cos \varphi& \frac{\vt}{v} \sin \varphi - \frac{\vr}{v} \cos \varphi&0
\end{array} \right]
\left[\hspace{-2mm}\begin{array}{l}
\displaystyle{\mathbf{e}_X }\\
\displaystyle{\mathbf{e}_Y} \\
\displaystyle{ \mathbf{e}_Z}
\end{array}\hspace{-2mm}\right],
\end{equation}
with $\vr$ (resp. $\vt$) the radial (resp. tangential) component of the test  star's velocity.
We can finally relate the frame~$(X,Y,Z)$ to the frame~$(x,y,z)$ through the relations
\begin{align}
\label{eq:xyz_to_XYZ}
\left[\hspace{-2mm}\begin{array}{l}
 \mathbf{e}_X \\
 \mathbf{e}_Y\\
\mathbf{e}_Z
\end{array}\hspace{-2mm}\right]
& =  \left[\begin{array}{ccc}
\cos I  & 0 &\sin I   \\
0&1&0 \\
-\sin I& 0&\cos I  
\end{array}\right]
 \left[\hspace{-2mm}\begin{array}{l}
\displaystyle{\mathbf{e}_x }\\
\displaystyle{\mathbf{e}_y} \\
\displaystyle{\mathbf{e}_z}
\end{array}\hspace{-2mm}\right].
\end{align}

We can now compute $L$, the norm of $\bL$.
As it is a norm,
we can choose any frame to compute it.
The frame $(1,2,3)$ is a convenient choice
as this frame is used to compute the integrands of equations~\eqref{eq:dvi}.
We write
\begin{align}
L^{\prime 2} &\!=\! (r_1 v_2'\!-\! r_2 v_1')^2 \!+\! (r_1 v_3' \!-\! r_3 v_1')^2 \!+\! (r_2 v_3' \!-\! r_3 v_2')^2,
\end{align}
where $(r_1,r_2,r_3)$ are the coordinates of $\br$ in the frame $(1,2,3)$.
Using equations~\eqref{eq:XYZ_to_123} and~\eqref{eq:xyz_to_XYZ},
we obtain
\begin{equation}
r_1 = \frac{\vr}{v} r \quad;\quad r_2 = 0 \quad ; \quad r_3 = \frac{\vt}{v} r.
\end{equation}
Therefore, one gets
\begin{align}
\label{eq:arg_Lp}
L^{\prime } &=r \sqrt{    v_2^{\prime 2} +  \bigg ( \frac{\vr}{v}  v_3' -  \frac{\vt}{v}  v_1'  \bigg)^2  }
\\
&= r\sqrt{   (w \sin \vartheta \cos \phi)   ^2 +  \bigg (\hspace{-0.5mm}\vt+ \frac{\vr}{v}  w \sin \vartheta \sin \phi-  \frac{\vt}{v}    w \cos \vartheta  \bigg)^2},\notag
\end{align}
given that $\bv'\!=\!\bv-\bw$.

Let us finally compute $\Lz'$.
In the $(x,y,z)$ frame, it reads
\begin{equation}
\label{eq:Lzp}
\Lz' = x v_y' - y v_x'.
\end{equation}
From equation~\eqref{eq:xyz_to_XYZ},
the positions read
\begin{subequations}
\label{eq:xyz_XYZ}
\begin{align}
x &= X \cos I = r \cos \varphi \cos I,\\
y &= Y = r \sin\varphi ,\\
z &= X \sin I = r \cos\varphi \sin I ,
\end{align}
\end{subequations}
using ${X\!=\!r \cos \varphi}$ and ${Y\!=\!r \sin\varphi}$.
We then obtain the velocities
\begin{subequations}
\begin{align}
v_x &=  \vr \cos \varphi \cos I-\vt  \sin \varphi \cos I,\\
v_y &=  \vr \sin \varphi + \vt \cos\varphi ,\\
v_z &=  \vr \cos \varphi\sin I-\vt  \sin \varphi \sin I, \label{eq:vz}
\end{align}
\end{subequations}
by taking the time derivative of equations~\eqref{eq:xyz_XYZ}. Let us now compute the background velocity components. First, using the left panel of Fig.~\ref{fig:coordinate_system}, we have
\begin{align}
\label{eq:vXYZ_xyz}
\left[\hspace{-2mm}\begin{array}{l}
v'_x\\
v'_y\\
v'_z
\end{array}\hspace{-2mm}\right]
& =  \left[\begin{array}{ccc}
\cos I  & 0 &-\sin I   \\
0&1&0 \\
\sin I& 0&\cos I  
\end{array}\right] 
 \left[\hspace{-2mm}\begin{array}{l}
\displaystyle{v'_X }\\
\displaystyle{v'_Y} \\
\displaystyle{v'_Z}
\end{array}\hspace{-2mm}\right].
\end{align}
Second, using the relation in equations~\eqref{eq:XYZ_to_123} (Fig.~\ref{fig:Coord_syst_1_spe} and right panel of Fig.~\ref{fig:coordinate_system}), we have  
\begin{equation}
\left[\hspace{-2mm}\begin{array}{l}
v'_X\\
v'_Y \\
v'_Z
\end{array}\hspace{-2mm}\right]=  \left[  \hspace{-2mm} \begin{array}{ccc}
\frac{\vr}{v} \cos\varphi- \frac{\vt}{v} \sin\varphi  & 0 &\frac{\vr}{v} \sin \varphi + \frac{\vt}{v}\cos \varphi   \\
\frac{\vr}{v} \sin\varphi+ \frac{\vt}{v}\cos \varphi&0& \frac{\vt}{v} \sin\varphi - \frac{\vr}{v} \cos \varphi  \\
0&1&0
\end{array}\hspace{-2mm} \right]
\left[\hspace{-2mm}\begin{array}{l}
\displaystyle{v'_1 }\\
\displaystyle{v'_2} \\
\displaystyle{v'_3}
\end{array}\hspace{-2mm}\right].
\end{equation}
Overall, we can rewrite equation~\eqref{eq:Lzp} as
\begin{align}
\Lz'  &= r \, \bigg(\frac{v'_1 \vt}{v}  - \frac{v'_3 \vr}{v}\bigg) \cos I  +r v'_2 \sin \varphi \sin I
\nonumber
\\
& = r \, \bigg(\vt - \frac{ \vt}{v} w\cos \vartheta + \frac{\vr}{v} w \sin \vartheta \sin \phi\bigg) \cos I
\nonumber
\\
&\quad - r w \sin \vartheta \cos \phi \sin \varphi \sin I .
\label{eq:arg_Lzp}
\end{align}
Finally, using the Cauchy--Schwarz inequality,
one can check that
\begin{align}
|\Lz'| \leq L' \sqrt{ \cos^2 I +  \sin^2 \varphi\sin^2 I},
\end{align}
which ensures that ${|\Lz'| \!\leq \!L'}$.

Equations~\eqref{eq:arg_Ep},~\eqref{eq:arg_Lp} and~\eqref{eq:arg_Lzp}
are the main results of this section.
They provide us with explicit expressions
for the arguments at which to evaluate ${ \Frot \!=\! \Frot (E' , L', \Lz') }$
in equation~\eqref{eq:dv_loc_ref}.

\section{3D diffusion coefficients}
\label{sec:Lz_coeffs}

The local in-plane diffusion coefficients,
that is, the diffusion coefficients in $E$ and $L$,
are given in appendix~{C} of \citet{BarOr2016}.
We reproduce them here for completeness.
They read
\begin{subequations}
\label{eq:nonrotatingdE}
\begin{align}
\dE &= \half \dvParSq +  \half \dvPerpSq + v    \dvPar, \\
\dESq &= v^2 \dvParSq , \\
\dL &= r \frac{\vt}{v} \dvPar + \frac{r^2}{4L} \dvPerpSq,\\
\dLSq &=r^2 \frac{\vt^2}{v^2} \dvParSq + \frac{r^2}{2}   \frac{\vr^2}{v^2} \dvPerpSq,\\
\dEL &= L \dvParSq.
\end{align}
\end{subequations}
In order to compute the 3D \FP\ equation~\eqref{eq:def_FP},
we also need the local diffusion coefficients involving ${ \Lz \!=\! \bL \!\cdot\! \mathbf{e}_{z} }$ .
Performing a first-order variation of ${\bL \!=\! \br \times \bv}$,
we can write
\begin{subequations}
\begin{align}
\Delta \bL &= \bigg(\frac{\Delta v_{\parallel}}{v} - \frac{\vr}{\vt} \frac{\Delta v_{3}}{v}\bigg) \, \bL + L \frac{\Delta v_{2}}{v} \bigg(\frac{\vr}{\vt} \mathbf{e}_3 - \hbv\bigg) ,
\\
 \mathbf{e}_3 & = \frac{\bv \times \bL}{|\bv \times \bL|} = \frac{v \hr - \vr \hbv}{\vt},
\end{align}
\end{subequations}
where $( \mathbf{e}_1, \mathbf{e}_2, \mathbf{e}_3)$ is the frame illustrated
in Fig.~\ref{fig:Coord_syst_1_spe}.
We can then write
\begin{equation}
 \mathbf{e}_3 \cdot  \mathbf{e}_z  = \sin I \, \bigg(\frac{\vr}{v}\sin\varphi+ \frac{\vt}{v} \cos \varphi \bigg) .
\end{equation}
As a result, it follows that
\begin{equation}
\frac{\vr }{v} \, \mathbf{e}_3 \cdot  \mathbf{e}_z - \frac{v_z \vt}{v^2} = \sin\varphi \sin I,
\end{equation}
with $\varphi$ given in Fig.~\eqref{fig:coordinate_system} and $v_z$ by equation~\eqref{eq:vz}.
Overall, we finally have
\begin{subequations}
\label{eq:coeffs_dELLz}
\begin{align}
\dLz &=  \frac{\Lz}{v}  \dvPar   ,\\
\dLzSq &= \bigg( \frac{\Lz}{L} \bigg)^2   \bigg(  \frac{L^2}{v^2} \dvParSq\!+ \! \frac{1}{2} \frac{r^2\vr^2 }{ v^2}  \dvPerpSq \bigg) \\
&+\frac{ r^2 \sin^2 \varphi}{2} \bigg(1\!-\!\frac{\Lz^2}{L^2}\bigg)     \dvPerpSq ,  \notag\\
\dELz &=  \Lz \dvParSq , \\
\dLLz  
&=  \frac{\Lz}{L}\bigg(\frac{L^2}{v^2}  \dvParSq  + \frac{1}{2}  \frac{ r^2 \vr^2}{ v^2}  \dvPerpSq \bigg) .
\end{align}
\end{subequations}
Equations~\eqref{eq:nonrotatingdE}
and~\eqref{eq:coeffs_dELLz} are the main results of this section.
They allow us to compute the local diffusion coefficients
in ${ (E , L , \Lz) }$ from the local velocity diffusion coefficients
given by equation~\eqref{eq:dv_loc_ref}.

\section{Orbit average}
\label{app:orbit_average}

The computation of the global diffusion coefficients 
in the \FP\ equation~\eqref{eq:def_FP}
requires that we orbit average the local diffusion coefficients
over the mean field orbit of the test star.
Since the background cluster is rotating,
this average involves a radial integration, but also an angular one.
This is what we detail here.

\subsection{Angular orbit average}
\label{app:orbit_average_angle}

We start from the generic formula of equation~\eqref{eq:generic_orbit_average}.
There, the average with respect to $\varphi$ must be performed with care
since $\Lz'$ depends on $\varphi$ (equation~\ref{eq:arg_Lzp}).
Fortunately, taking advantage of the particular structure
of the diffusion coefficients in $E$, $L$ and $\Lz$ (see equations~\ref{eq:nonrotatingdE} and~\ref{eq:coeffs_dELLz}),
we only need to compute the following four integrals over $\varphi$
\begin{align}
&\int \frac{\rd \varphi}{2\pi}  \dvPar ,\quad \int \frac{\rd \varphi}{2\pi}  \dvParSq ,
\nonumber
\\
&\int \frac{\rd \varphi}{2\pi}  \dvPerpSq ,\quad \int \frac{\rd \varphi}{2\pi}  \cos  2\varphi \, \dvPerpSq,
\label{eq:avg_theta_v}
\end{align}
where we used  ${ \sin^2 \varphi \!=\! \frac{1}{2}(1 \!-\! \cos 2\varphi) }$ for convenience.

In equation~\eqref{eq:dv_loc_ref},
the only dependence with respect to $\varphi$ is in the dependence
with respect to $\Lz'$ in $\Frot$.
In the particular case of the \LBD\ (equation~\ref{eq:LBD}),
we can perform explicitly
the average over $\varphi$.
Computing the integrals~\eqref{eq:avg_theta_v}
only requires the evaluation of the two  non-trivial integrals  
\begin{subequations}
\begin{align}
\label{eq:A1A2def}
A_1&=\int_{0}^{2\pi} \frac{\rd \varphi}{2\pi} \, \sgn(\Lz' /L') ,\\
 A_2&=\int_{0}^{2\pi} \frac{\rd \varphi}{2\pi} \, \cos 2 \varphi \, \sgn(\Lz'/L') ,
\end{align}
\end{subequations}
with $\Lz'$ following from equation~\eqref{eq:arg_Lzp}.
Introducing ${ \zeta\!=\!-(v_1' \vt/v \!-\! v_3' \vr/v) \cos I}$ and  ${\mu\! =\!  \zeta/( v_2' \sin I)}$,
these two integrals can be explicitly computed.
Their expressions are gathered in Table~\ref{tab:angleAvgA}.
\begin{table}
\centering
\setlength{\tabcolsep}{.5em}
 \begin{tabular*}{0.804\columnwidth}{@{}|c|c|c|c|}
\hline
$v_2'$ & $\mu$ & $A_1$ & $A_2$
\\[0.5ex]
\hline
0 & $\mathrm{undefined}$  &  $-\sgn(r \zeta )$  & 0
\\
$\neq0$ & $>1$ & $-1$  & 0
\\
$\neq0$ & $[-1,1]$ & $- 2\sin^{-1}\mu /\pi$  &$ -2 \mu \sqrt{1-\mu^2}/\pi $
\\
$\neq0$ & $<-1$ & $1$  & 0
\\
\hline
\end{tabular*}
\caption{Explicit values of $A_1$ and $A_2$ (equation~\ref{eq:A1A2def})
for the \LBD\ \DF\ (equation~\ref{eq:LBD}). 
}
\label{tab:angleAvgA}
\end{table}
This is the main result of this section.

From the numerical point of view,
this analytical calculation allows us to speed up
the computation of the \NR\ predictions,
as the angular integral is removed.
In addition, it also explicitly deals with the discontinuity
of the \LBD\ \DF\@: this enhances the numerical stability.

\subsection{Radial orbit average}
\label{app:orbit_average_radial}

Having performed the angle average in Appendix~\ref{app:orbit_average_angle}, equation~\eqref{eq:generic_orbit_average} takes the form 
\begin{equation}
\label{eq:radial_avg}
D_{X} = \frac{\Omega_r}{\pi} \int_{\rrp}^{\rra} \frac{\rd r}{|\vr|}  \langle \Delta X \rangle_{\varphi} (r) ,
\end{equation}
where $\langle \cdot  \rangle_{\varphi}$ stands for the angle average.
As described in appendix~{F2} of~\citetalias{Tep2022},
equation~\eqref{eq:radial_avg} can be efficiently performed
by changing variable from the radius to an effective anomaly.
In particular, this removes the integrable singularity $1/|\vr|$ 
at pericentre and apocentre.
We follow the exact same approach here,
and perform the radial average using 50 sampling nodes.

\section{Fokker--Planck and coordinates}
\label{app:action_coeffs}

Depending on the quantities we wish to investigate,
we may wish to change of coordinates system.
Fortunately, it is possible to transform one \FP\ description into another
under a change of variable.
We describe a few relevant examples in this appendix.

\subsection{Generic change of coordinates}
\label{app:convert_FP_generic}

Given some coordinates $\bx$, the \FP\ equation generically reads \citep{Risken1996}
\begin{align}
\frac{\p F(\bx,t)}{\p t}  &= - \frac{\p }{\p \bx} \cdot \mbF_{\bx} (\bx)\\
 &= - \frac{\p }{\p \bx} \cdot \bigg[ \bD_{\bx} (\bx) \, F(\bx)- \frac{1}{2} \frac{\p }{\p \bx} \!\cdot\! \bigg( \bD_{\bx\bx} (\bx) \, F(\bx) \bigg) \bigg] ,\notag
\label{eq:generic_FP_eq}
\end{align}
with the diffusion coefficients
\begin{equation}
\bD_{\bx} (\bx) \!=\!
\begin{bmatrix}
D_{x_1}
\\
D_{x_2}
\\
D_{x_3}
\end{bmatrix}
; \quad
\bD_{\bx\bx} (\bx) \!=\!
\begin{bmatrix}
D_{x_1x_1} \!\!&\!\! D_{x_1 x_2}  \!\!&\!\! D_{x_1 x_3}
\\
D_{x_1 x_2}   \!\!&\!\! D_{x_2x_2}  \!\!&\!\! D_{x_2 x_3}
\\
D_{x_1 x_3}   \!\!&\!\!D_{x_2 x_3}  \!\!&\!\! D_{x_3x_3}
\end{bmatrix} ,
\label{def_Dx_Dxx}
\end{equation}
and the DF in $\bx$ space, ${F(\bx)}$.
In our case, depending on the context,
${\bx\!=\!(x_1,x_2,x_3)}$ may either stand for $(E,L,\Lz)$,  $(\Jr,L,\Lz)$ or  $(\Jr,L,\cos I)$.

Following \citet{Risken1996} and \citet{BarOr2016},
one can easily rewrite the \FP\ equation
within some new coordinates ${ \bx' (\bx) }$.
The new diffusion coefficients read
\begin{subequations}
\label{eq:change_var_DiffCoeffs}
\begin{align}
D_{x_l'} & = \sum_{k} \frac{\partial x'_{l}}{\partial x_{k}}D_{x_k} + \frac{1}{2} \sum_{k, r}\frac{\partial^{2}x'_{l}}{\partial x_{r}\partial x_{k}}D_{x_r x_k} ,
\\
 D_{x_l' x_m'} & = \sum_{k, r} \frac{\partial x'_{l}}{\partial x_{r}}\frac{\partial x'_{m}}{\partial x_{k}}D_{x_r x_k} .
\end{align}
\end{subequations}
These coefficients source a \FP\ equation in $\bx'$-space
for the \DF, ${F'(\bx')}$,
reading
\begin{equation}
F'(\bx') = \bigg| \frac{\p \bx}{\p \bx'} \bigg| \,F(\bx),
\label{eq:new_DF_FP}
\end{equation}
with $|\p \bx/\p \bx'|$ the inverse Jacobian
of the coordinate transform.

\subsection{From $(E,L,\Lz)$ to $(\Jr,L,\Lz)$}
\label{app:convert_E_to_Jr}

Following section~\ref{sec:NR},
we have at our disposal diffusion coefficients
within the coordinates ${(E,L,\Lz)}$.
Owing to equations~\eqref{eq:change_var_DiffCoeffs},
the diffusion coefficients in the action space $(\Jr[E,L],L,\Lz)$
are easily computed.
This is already detailed in appendix~{F3} of \citetalias{Tep2022}.
Ultimately, we have at our disposal
the diffusion coefficients
\begin{align}
\bD_{1} (\bJ)\!=\!
\left[\hspace{-2mm}\begin{array}{l}
D_{\Jr}\\
D_{L}\\
D_{\Lz}
\end{array}\hspace{-2mm}\right]
; \quad
\bD_{2} (\bJ) \!=\!
 \left[\hspace{-2mm}\begin{array}{ccc}
D_{\Jr\Jr}  \!\!\!&\!\!\! D_{\Jr L}    \!\!\!&\!\!\!D_{\Jr \Lz}   \\
D_{\Jr L}  \!\!\!&\!\!\! D_{L L}  \!\!\!&\!\!\!D_{L\Lz} \\
D_{\Jr \Lz}  \!\!\!&\!\!\! D_{L\Lz}  \!\!\!&\!\!\!D_{\Lz\Lz}  
\end{array}\hspace{-2mm}\right]\! ,\!
\end{align}
which source the \FP\ evolution of  $\Frot(\Jr,L,\Lz)$,
the cluster's \DF\ in $(\Jr,L,\Lz)$.

\subsection{From $(\Jr,L,\Lz)$ to $(\Jr,L,\cos I)$}
\label{app:convert_Lz_to_cosI}

Let us now change of coordinates
from ${\bJ \!=\! (\Jr,L,\Lz)}$ to ${\bJc \!=\! (\Jr,L,\cos I\!=\!\Lz/L)}$.
Applying equations~\eqref{eq:change_var_DiffCoeffs}, 
we obtain the new diffusion coefficients 
\begin{subequations}
 \label{eq:change_var_Lz_cosI}
\begin{align}
  D_{\cosI} &  = - \frac{\cos I \, D_{L}}{L}  \!+\! \frac{D_{\Lz}}{L} \!+\! \frac{\cos I \, D_{L L}}{L^2} \!-\! \frac{D_{L \Lz}}{L^{2}} ,
  \\
  D_{\Jr \cosI} & = -\frac{\cos I \, D_{\Jr L}}{L} \! + \!\frac{D_{\Jr \Lz}}{L} = 0,
  \\
   D_{L \cosI} & = -\frac{\cos I \, D_{L L}}{L} \!+\! \frac{D_{L \Lz}}{L} = 0,
   \\
   D_{ \cosI \!\cosI} & = \frac{\cos^2 I \,  D_{L L}}{L^2} \!-\! \frac{2 \cos I \, D_{L \Lz}}{L^2} \!+\! \frac{D_{\Lz \Lz}}{L^2} ,
 \end{align}
\end{subequations}
while the other coefficients stay unchanged.
Importantly, we point out that ${ D_{\Jr \cos I} \!=\! D_{L \cos I} \!=\! 0 }$.
This comes from the relations
${D_{E\Lz} \!=\! \cos I \, D_{E L}}$ and  ${D_{L\Lz} \!=\! \cos I \, D_{L L}}$,
which are inferred from equations~\eqref{eq:nonrotatingdE} and~\eqref{eq:coeffs_dELLz}.
Hence, we are left with the diffusion coefficients
\begin{align}
\label{eq:D_JrLcosI}
\bD_{1} (\bJc)\!=\!
\left[\hspace{-2mm}\begin{array}{l}
D_{\Jr}\\
D_{L}\\
D_{\cosI}
\end{array}\hspace{-2mm}\right]
; \quad
\bD_{2} (\bJc) \!=\!
 \left[\hspace{-2mm}\begin{array}{ccc}
D_{\Jr\Jr}  \!\!\!&\!\!\! D_{\Jr L}    \!\!\!&\!\!\!0   \\
D_{\Jr L}  \!\!\!&\!\!\! D_{L L}  \!\!\!&\!\!\!0 \\
0  \!\!\!&\!\!\! 0  \!\!\!&\!\!\!D_{\cosI\!\cosI}  
\end{array}\hspace{-2mm}\right]\! .\!
\end{align}
Overall, they drive the \FP\ evolution
of ${F(\bJc) \!=\! L \Frot(\bJ)}$,
the cluster's \DF\ in ${ (\Jr , L , \cos I) }$.

\section{$N$-body simulations}
\label{sec:NBODY}

The simulations presented throughout the main text
were performed using the direct $N$-body code
{\sc \small NBODY6++GPU}~\citep{Wang2015},
version 4.1.
We used the exact same run parameters as in appendix~{G} of~\citetalias{Tep2022},
and drew the initial conditions using {\sc PlummerPlus.py}~\citep{Breen2017}.
We also refer to the aforementioned appendix for the $N$-body measurements, 
and to Table~\ref{table:para_NBODY}
for all our binning parameters.
\begin{table}
\centering
\setlength{\tabcolsep}{.5em}
 \begin{tabular*}{0.68\columnwidth}{@{}lccc}
\hline
\hline
$q$ & 1 & 0  & -6   \\
\hline
$\small \Nrun$ & \small 50 & \small 50 & \small 50  \\
$\small \tlast \, [\rHU]$& \small 1000 & \small 1000 & \small 1000    \\
$\small (\Jr^{\min} , \Jr^{\max})$ &\small (0, 0.55) &\small (0, 0.55) &\small (0, 0.5)  \\
$(\small L^{\min} , L^{\max}) $&\small (0,1.05) &\small (0,1.05) & \small (0,1.1)  \\
$\small (N_{\Jr} , N_{L}) $ & \small (20,20) &\small (20,20) &\small (30,40)   \\
$\small (N_{\Jr} , N_{\cos I}) $ &\small (20,20) &\small (20,20) &\small (20,20)   \\
$\small (N_{L} , N_{\cos I}) $ &\small (20,20) & \small (20,20) &\small (20,20)   \\
\hline
\end{tabular*}
\caption{Detailed parameters for the measurements in $N$-body simulations,
following the same notation as in appendix~{G1} of~\citetalias{Tep2022}.
To measure relaxation rate in Fig.~\ref{fig:dFdt_NBODY_JrL},
we bin the ${(\Jr,L)}$ domain
in ${ N_{\Jr} \!\times\! N_{L} }$ uniform bins
within the region ${\Jr^{\min}\! \leq \!\Jr \!\leq \!\Jr^{\max}}$
(similarly for $L$).
We use a similar approach for ${(\Jr,\cos I)}$ and ${(L,\cos I)}$.
All quantities are in physical units ${ G\!=\!M\!=\!b\!=\!1 }$, if not stated otherwise. 
}
\label{table:para_NBODY}
\end{table}

Each $N$-body realisation was composed of ${N\!=\! 10^5}$ (resp.\ ${N\!=\! 10^4}$) stars and integrated up to ${t_{\max}\! =\! 1 000\,\rHU}$ (resp.\ ${t_{\max}\! =\! 4 000\,\rHU}$)
with a dump every ${1 \rHU}$.
On a 40-core CPU node with a single V100 GPU,
one simulation typically required ${\sim \!24}$ h (resp. ${\sim \!12}$ h) of computation.
Ensemble averages were performed over 50 independent runs.

In practice, the main difficulties in the $N$-body measurements are
(i) the estimation of the instantaneous potential -- necessary to compute
the instantaneous radial action, $\Jr$;
(ii) the determination of the bins' size in action space;
(iii) the estimation of the relaxation rate $\p F/\p t$ via finite differences.
This is especially true for highly tangentially anisotropic clusters (e.g., ${q\!=\!-6}$),
where stars are closely stacked near the ${ \Jr \!=\! 0 }$ axis.
For these three difficulties,
we use the same approach as in~\citetalias{Tep2022}.

\section{Sphericity}
\label{app:sphericality}

In order to track the clusters' sphericity in Fig.~\ref{fig:sphericality_rotation},
we introduce the 3D inertia-like matrix
\begin{equation}
\mbI = \sum_{k = 1}^{N} \rho_{k}^{2} \bigg( \br_k^{\rT} \br_{k} \bI - \br_{k} \br_{k}^{\rT} \bigg) \bigg/ \sum_{k=1}^{N} \rho_{k}^{2} ,
\label{eq:def_I}
\end{equation}
with $\br_{k}$ the location of the $k$-th particle,
$\rho_{k}$ its local density~\citep[see][]{Casertano1985}
and $\bI$ the 3D identity matrix.
In that definition, the extra $\rho_{k}^{2}$ factors
enhance the contributions from the regions close to the centre.

The matrix $\mbI$ is obviously symmetric.
It is also semi-definite positive since
for any $\by \in \mathbb{R}^{3}$, we have
\begin{align}
\by^{\rT} \mbI \, \by & = \sum_{k} \rho_{k}^{2} \, \bigg( \br_{k}^{\rT} \br_{k} \by^{\rT} \by - \by^{\rT} \br_{k} \br_{k}^{\rT} \by \bigg)\bigg/ \sum_{k} \rho_{k}^{2}
\nonumber
\\
&= \sum_{k} \rho_{k}^{2} \, \bigg( |\br_{k}|^{2} \, |\by_{2}|^{2} \!-\! | \br_{k} \!\cdot\! \by |^{2} \bigg)  \bigg/ \sum_{k} \rho_{k}^{2} \geq 0 ,
\end{align}
following Cauchy--Schwarz's inequality.
As a result, $\mbI$ has three positive eigenvalues,
${ \{ \lambda_{i} \}_{i} }$,
which encapsulate the cluster's sphericity.
Indeed, spherically symmetric clusters
have all their eigenvalues equal.

We generically define the cluster's sphericity via
${h \!=\! \lambda_{\min} / \lambda_{\max}}$,
which we estimated from $N$-body simulations.
To reduce shot noise in that measurement,
we averaged $h$ over realisations
 as follows.
First, we computed the elementary symmetric polynomials
${ \alpha \!=\! \lambda_1 \!+\! \lambda_2 \!+\! \lambda_3 }$,
${ \beta \!=\! \lambda_1 \lambda_2 \!+\! \lambda_1 \lambda_3 \!+\! \lambda_2 \lambda_3 }$
and ${ \gamma \!=\! \lambda_1 \lambda_2 \lambda_3 }$
for every cluster and every timestep.
We then averaged the values of ${ (\alpha , \beta , \gamma) }$ over all realisations.
From these values, we estimated the eigenvalues $\lambda_i$
as the three (positive) roots of the polynomial
${\lambda^3 \!-\! \langle\alpha\rangle \lambda^2 \!+\! \langle\beta\rangle \lambda \!-\! \langle\gamma\rangle}$.\footnote{The positivity of $\alpha,\beta,\gamma$ -- and hence of their averages -- ensures the positivity of these roots.}

\begin{figure*}
\centering
\includegraphics[width=0.225 \textwidth]{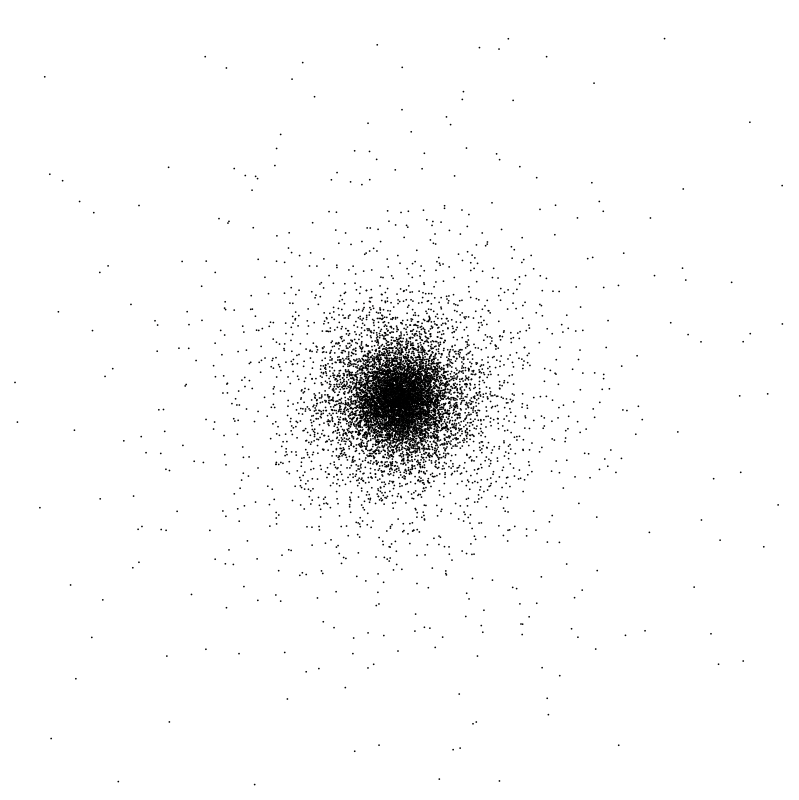}
\includegraphics[width=0.225 \textwidth]{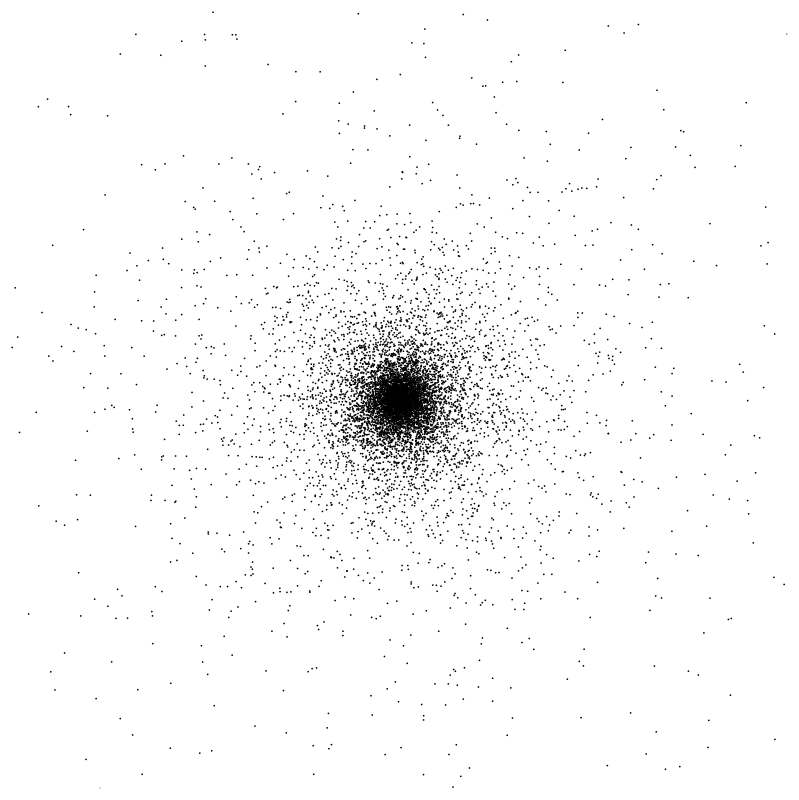}    
\includegraphics[width=0.225 \textwidth]{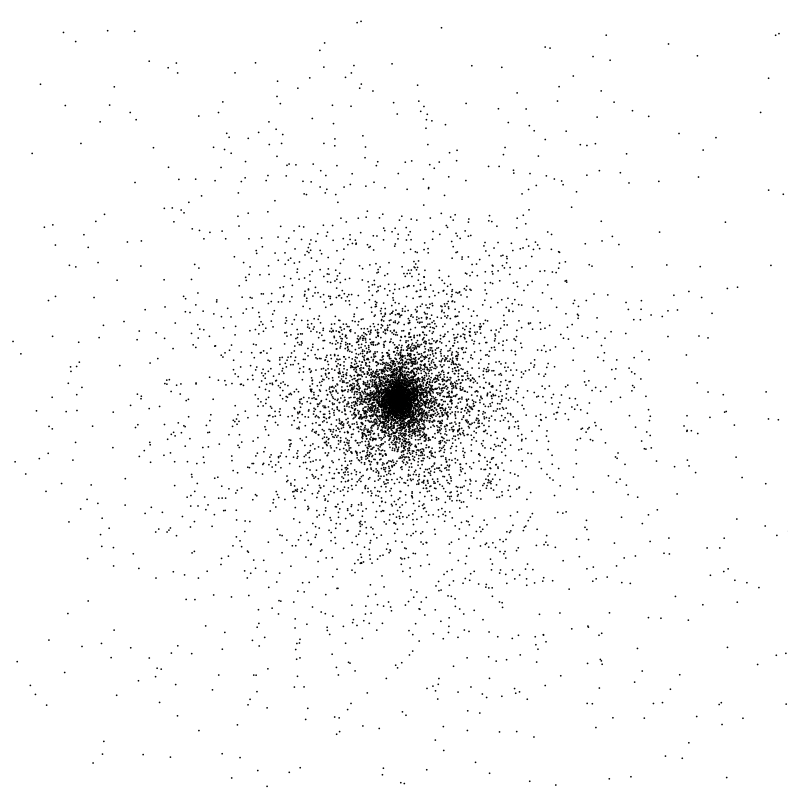}
\includegraphics[width=0.225 \textwidth]{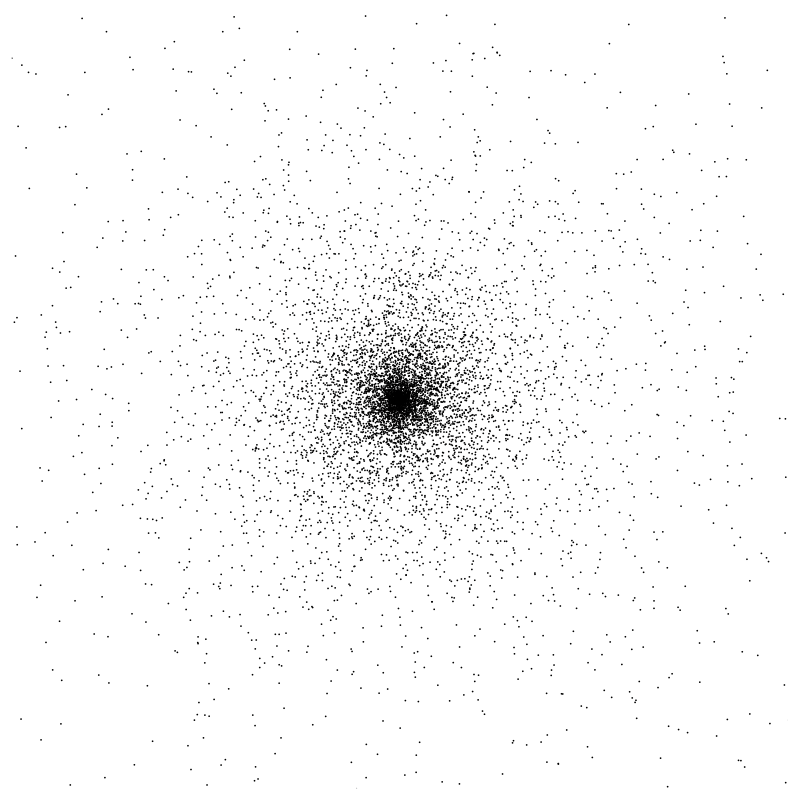}
\caption{Snapshots of the distribution of stars in a rotating (${ \alpha \!=\! 0.25 }$)
isotropic (${ q \!=\! 0 }$) Plummer cluster with ${N\!=\!10^4}$ stars,
projected on the rotation plane.
From left to right, this corresponds to ${t\!=\!0,1000,2000,3000\, \rHU}$.
As time evolves, the cluster contracts and the central density increases.
Notwithstanding, the cluster retains its initial spherical symmetry.
}
\label{fig:snapshots_nbody}
\end{figure*}

In Fig.~\ref{fig:sphericality_rotation}, we illustrate
the evolution of the sphericity, $h$, for various amounts
of anisotropy and rotation.
We typically find $ {h\! \simeq\! 0.996}$,
that is, clusters remain reasonably spherically symmetric
throughout their evolution.
As as visual check of this conclusion, we represent in Fig.~\ref{fig:snapshots_nbody}
the late time stellar distribution of a rotating isotropic globular cluster with ${N\!=\!10^4}$.
Even close to core collapse, the cluster remains spherically symmetric.

\section{Rotating King model}
\label{app:King}

In section~\ref{sec:gravo_gyro}, we mentioned the gravo-gyro catastrophe,
introduced in \citet{Hachisu1979} as the rotational counterpart
of the gravothermal catastrophe \citep{LyndenBell1968}
to explain an apparent acceleration of core collapse with rotation.
To study this phenomenon, \cite{Einsel1999}
considered the secular evolution of a rotating King model with the initial \DF\
\begin{equation}
\label{eq:def_DF_King}
\Frot(E,\Lz) \propto  (\re^{ - \beta E}-1)\  \re^{-\beta \Omega_0 \Lz},
\end{equation}
where ${\Frot\!=\!\Frot(\br,\bv)\!=\!\Frot(E,\Lz)}$ is the DF in ${(\br,\bv)}$.
In this \DF\@, ${\beta\!=\!1/\sigma_{\rc}^2}$ is an inverse temperature
with $\sigma_{\rc}$ the central velocity dispersion,
while ${\omega_0\!=\!\sqrt{9/(4\pi G n_{\rc})} \, \Omega_0}$ is a rotation parameter with $n_{\rc}$ the central density and $\Omega_0$ the angular velocity in the cluster's centre \citep{Lagoute1996}.
In particular, $\beta$ is related to the so-called King parameter \citep{King1966},
defined by ${W_0\!=\!-\beta (\psi-\psi_{\rt})}$
where $\psi_{\rt}$ is the potential at the cluster's edge.
The rotating King models from~\cite{Einsel1999}
can therefore be parametrised by $(W_0,\omega_0)$.
In practice, \cite{Einsel1999} integrated these models up
to core collapse using a \FP\ scheme in ${(E , \Lz)}$
for different rotation parameters, $\omega_0$.
This is presented in their fig.~{2}.
In particular, they demonstrated a clear correlation between the time of core collapse
and the rotation parameter of the King model.
The faster a cluster rotates,
the shorter the core collapse time.
Importantly, as emphasised in section~\ref{sec:gravo_gyro},
we recall that as one varies $\omega_{0}$,
the cluster's mean potential also varies.

\section{2D Fokker-Planck equations}
\label{app:2D_FP}

In practice, it is convenient to study long-term relaxation
through two-dimensional projections of action space.
To do so, we must integrate over one coordinate
the 3D \FP\ equation
expressed either in ${ (\Jr , L , \Lz) }$ (appendix~\ref{app:convert_E_to_Jr})
or in ${ (\Jr , L , \cos I) }$ (appendix~\ref{app:convert_Lz_to_cosI}).
In this appendix, we compute the 2D \FP\ equations
in ${ (\Jr , L) }$, ${ (\Jr , \cos I) }$ and ${ (L , \cos I) }$
that respectively drive the evolution of the \DFs\
\begin{subequations}
\begin{align}
F(\Jr,L) &= \int_{-L}^{L} \rd \Lz \,\Frot(\Jr,L,\Lz) ,
\\
F(\Jr,\cos I) &= \int_{0}^{+ \infty} \rd L \,F(\Jr,L,\cos I),
\\
F(L,\cos I) &= \int_{0}^{+ \infty} \rd \Jr \,F(\Jr,L,\cos I).
\end{align}
\end{subequations}

\subsection{Equation in $(\Jr,L)$}
\label{app:FP_2d_JrL}

We start from equation~\eqref{eq:def_FP}, which describes diffusion in ${ (\Jr , L , \Lz) }$,
and write the flux divergence as
\begin{align}
  \frac{\p \Frot}{\p t} &= -\bigg(\frac{\p \mF_{\Jr}}{\p \Jr} +\frac{\p \mF_{L}}{\p L} +\frac{\p \mF_{\Lz}}{\p \Lz} \bigg) .
  \end{align}
Integrating over $\Lz$ yields
\begin{subequations}
\label{eq:integrated_flux_Lz}
\begin{align}
\!\! \int_{-L}^{L} \!\!\!\! \rd \Lz\,\frac{\p \mF_{\Jr}}{\p \Jr} &= \frac{\p }{\p \Jr} \!\! \int_{-L}^{L} \!\!\!\! \rd \Lz \, \mF_{\Jr} ,
\\
\!\! \int_{-L}^{L} \!\!\!\! \rd \Lz\,\frac{\p \mF_{L}}{\p L}  & = - \mF_{L}(\Lz\!=\!L) \!-\! \mF_{L}(\Lz\!=\!-L)
 + \frac{\p }{\p L} \!\! \int_{-L}^{L} \!\!\!\! \rd \Lz \, \mF_{L} ,
\\
\!\! \int_{-L}^{L} \!\!\!\! \rd \Lz\,\frac{\p \mF_{\Lz}}{\p \Lz} &=  \mF_{\Lz}(\Lz\!=\!L) - \mF_{\Lz}(\Lz\!=\!-L).
  \end{align}
  \end{subequations}
In addition, because the flux cannot exit action space,
we also have ${\mF_{\Lz}(\Lz\!=\!\pm L) \!=\!\pm \mF_{L}(\Lz\!=\!\pm L)}$.
As a result, all the non-integral terms in equations~\eqref{eq:integrated_flux_Lz}
cancel one another.
We are then left with the 2D equation
\begin{equation}
\label{eq:FP_JrL}
\frac{\p F(\Jr,L,t) }{\p t}= - \frac{\p }{\p (\Jr, L)} \cdot \!\! \int_{-L}^{L} \!\! \rd \Lz \,
\begin{bmatrix}
\mF_{\Jr} (\Jr , L , \Lz)
\\[1.0ex]
\mF_{L} (\Jr , L , \Lz)
\end{bmatrix} .
\end{equation}
It is this rewriting that is used in section~\ref{sec:NR_pred}. To obtain Fig.~\ref{fig:dFdt_NR_JrL}, the $\Lz$-integrals in equations~\eqref{eq:integrated_flux_Lz} are sampled with 50 nodes. The ${(w,\vartheta,\phi)}$-integrals of equations~\eqref{eq:dv_loc_ref} are sampled with 100 $w$-nodes, 100 $\vartheta$-nodes and 200 $\phi$-nodes.

\subsection{Equation in $(\Jr,\cos I)$}
\label{app:FP_2d_LcosI}

We start from the diffusion coefficients in equation~\eqref{eq:D_JrLcosI}, which describe diffusion in ${ (\Jr , L , \cos I) }$,
for the \DF\@, ${F \!=\!L \Frot}$.
We write the flux divergence as
\begin{align}
\label{eq:FP_3d_cosI}
  \frac{\p F}{\p t} &= -\bigg(\frac{\p \mF_{\Jr}}{\p \Jr} +\frac{\p \mF_{L}}{\p L} +\frac{\p \mF_{\cosI}}{\p \cosI} \bigg).
  \end{align}
Integrating over $L$ yields
\begin{subequations}
\label{eq:integrated_flux_L}
\begin{align}
\!\! \int_{0}^{\infty} \!\!\!\! \rd L\,\frac{\p \mF_{\Jr}}{\p \Jr} &= \frac{\p }{\p \Jr} \!\! \int_{0}^{+ \infty} \!\!\!\! \rd L \, \mF_{\Jr} ,\\
\!\! \int_{0}^{\infty} \!\!\!\! \rd L\,\frac{\p \mF_{L}}{\p L}  &=  \mF_{L}(L\!=\! + \infty) - \mF_{L}(L\!=\!0), \label{eq:FL_integrated}\\
\!\! \int_{0}^{\infty} \!\!\!\! \rd L\,\frac{\p \mF_{\cosI}}{\p \cosI} &=  \frac{\p }{\p \cosI} \!\! \int_{0}^{+ \infty} \!\!\!\! \rd L \, \mF_{\cosI}.
  \end{align}
  \end{subequations}
Because the flux cannot leave action space, equation~\eqref{eq:FL_integrated} vanishes.
Then, equation~\eqref{eq:FP_3d_cosI} readily reduces to a 2D \FP\ equation in $(\Jr,\cos I)$,
which we use in section~\ref{sec:NR_pred_cosI}. To obtain Fig.~\ref{fig:dFdt_NR_JrCosI}, the $L$-integrals in equations~\eqref{eq:integrated_flux_L} are sampled with 50 nodes for ${ 0 \!\leq\! L \!\leq\! 3 \, \Lo }$, where ${\Lo\!=\!\sqrt{G M b}}$ is the typical action.
The ${(w,\vartheta,\phi)}$-integrals of equations~\eqref{eq:dv_loc_ref}
are sampled with 100 $w$-nodes, 100 $\vartheta$-nodes and 800 $\phi$-nodes.

\subsection{Equation in $(L,\cos I)$}
\label{app:FP_2d_LcosItrue}

As in appendix~\ref{app:FP_2d_LcosI}, we start from the diffusion coefficients in equation~\eqref{eq:D_JrLcosI}. Integrating equation~\eqref{eq:FP_3d_cosI} over $\Jr$ yields
\begin{subequations}
\label{eq:integrated_flux_Jr}
\begin{align}
\!\! \int_{0}^{\infty} \!\!\!\! \rd \Jr\,\frac{\p \mF_{\Jr}}{\p \Jr} &=  \mF_{\Jr}(\Jr\!=\! + \infty) - \mF_{\Jr}(\Jr\!=\!0), \label{eq:FJr_integrated}\\
\!\! \int_{0}^{\infty} \!\!\!\! \rd \Jr\,\frac{\p \mF_{L}}{\p L}  &=\frac{\p }{\p L} \!\! \int_{0}^{+ \infty} \!\!\!\! \rd \Jr \, \mF_{L} \,, \\
\!\! \int_{0}^{\infty} \!\!\!\! \rd \Jr\,\frac{\p \mF_{\cosI}}{\p \cosI} &=  \frac{\p }{\p \cosI} \!\! \int_{0}^{+ \infty} \!\!\!\! \rd \Jr \, \mF_{\cosI}.
  \end{align}
  \end{subequations}
Once again, because the flux cannot leave action space,
equation~\eqref{eq:FJr_integrated} vanishes.
Then, equation~\eqref{eq:FP_3d_cosI} reduces to a 2D \FP\ equation in $(L,\cos I)$,
which we use in appendix~\ref{app:L_cosI_diff}. To obtain Fig.~\ref{fig:dFdt_NR_LCosI}, the $\Jr$-integrals in equations~\eqref{eq:integrated_flux_Jr} are sampled with 50 nodes for ${ 0 \!\leq\! \Jr \!\leq\! 10 \, \Lo }$.
The ${(w,\vartheta,\phi)}$-integrals of equations~\eqref{eq:dv_loc_ref} are sampled with 100 $w$-nodes, 100 $\vartheta$-nodes and 1600 $\phi$-nodes.

\section{Evaluating the discontinuity at $\cos I = 0$}
\label{app:disc_cosI_0}

\begin{figure} 
   \centering
   \includegraphics[width=0.49 \textwidth]{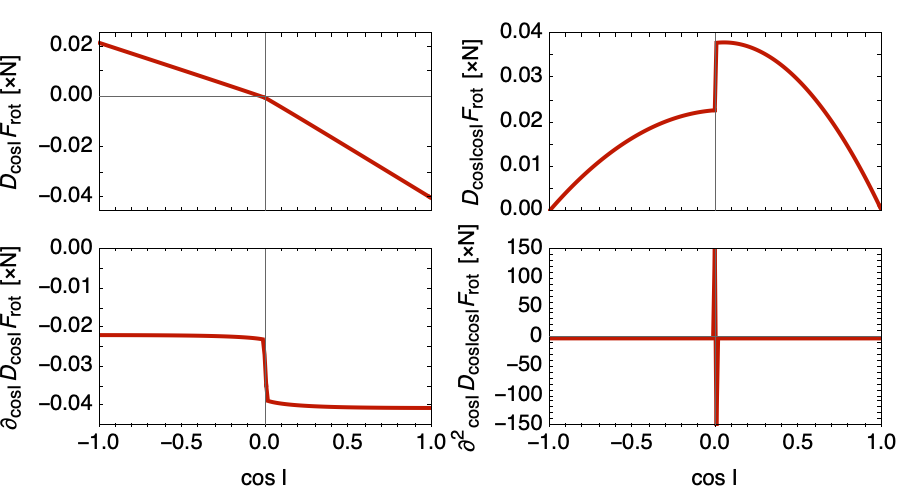}  
   \caption{Representation for a Plummer cluster $(q,\alpha)\!=\!(0,0.25)$ of the $\cos I$ terms, $D_{\cosI} \Frot$ (on the top left) and $D_{\cosI\cosI} \Frot$ (on the top right), respectively with the first (on the bottom left) and second (on the bottom right) derivative with respect to $\cos I$. In particular, the bottom right panel displays an approximate $\delta'(\cos I)$ behaviour -- the amplitude of the jumps tends to infinity as one reduces the step in the finite differentiation scheme -- which is the expected theoretical relaxation induced by the $\sgn$ discontinuity of $\Frot$ at $t=0$. 
      }
   \label{fig:disc_dirac_0}
 \end{figure}

In this appendix, we highlight the sharp discontinuity of the relaxation rate
that occurs at $\cos I\!=\!0$
as a result of the discontinuous \LBD\ distribution of orientations (equation~\ref{eq:LBD}).
This singularity has been removed from Figures~\ref{fig:dFdt_NR_JrCosI} and \ref{fig:dFdt_NR_LCosI}
(see appendices~\ref{app:FP_2d_LcosI} and~\ref{app:FP_2d_LcosItrue}).
Because equation~\eqref{eq:LBD} is discontinuous, we expect that taking the first and second derivatives with respect to $\cos I$ will yield a theoretical $\delta'(\cos I)$ behaviour around $\cos I\!=\!0$. To that aim, we show in Figure~\ref{fig:disc_dirac_0} the quantities $D_{\cosI} \Frot$ and $D_{\cosI\cosI} \Frot$, respectively with their first and second derivatives with respect to $\cos I$.
In particular, $\p^2 (D_{\cosI\cosI} \Frot)/\p \cosI^2$ has a  $\delta'(\cos I)$ component near $\cosI\!=\!0$. In practice, this is smoothed by the finite differentiation used here. Indeed, reducing the finite differentiation step sharpens the discontinuity to higher and higher values,
hence converging to the true ${ \delta'(\cos I) }$ behaviour.

\section{Relaxation in $(L,\cos I)$}
\label{app:L_cosI_diff}

\begin{figure*} 
  \centering
       \includegraphics[width=0.80 \textwidth]{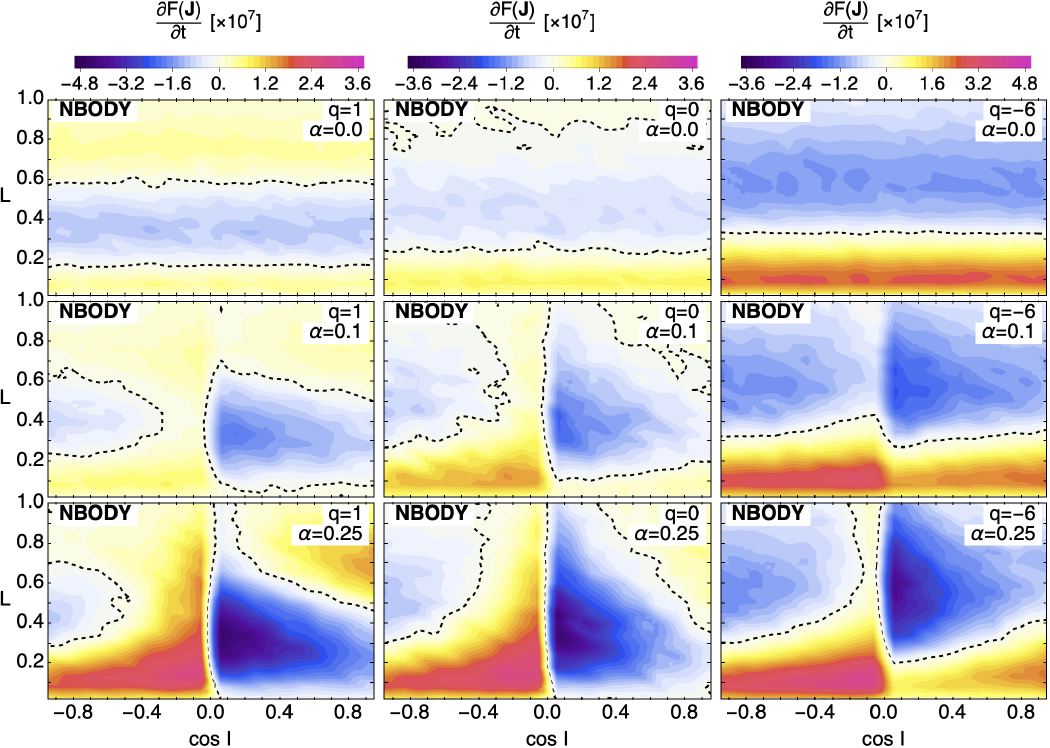}  
   \caption{Same as in Fig.~\ref{fig:dFdt_NBODY_JrCosI} but in $(L,\cos I)$.
   Diffusion reshuffles orbital inclinations
   toward a more affine distribution in $\cos I$.
   }
   \label{fig:dFdt_NBODY_LCosI}
 \end{figure*}

\begin{figure*} 
   \centering
   \includegraphics[width=0.80 \textwidth]{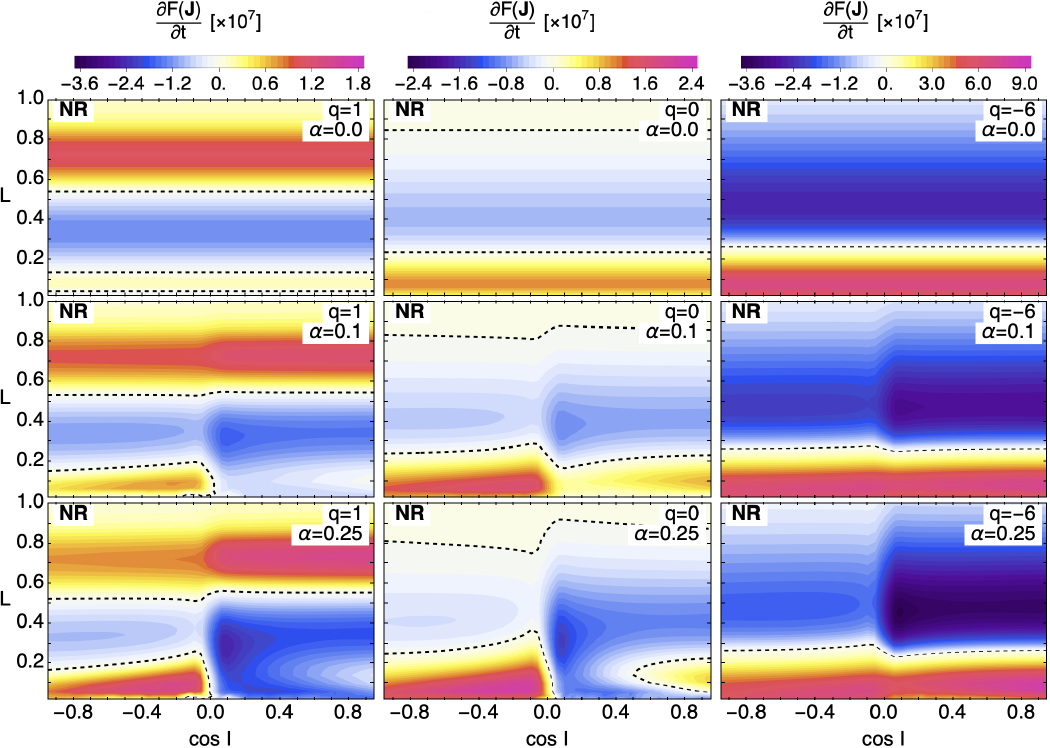}  
   \caption{Same as Fig.~\ref{fig:dFdt_NR_JrCosI} but in $(L,\cos I)$. 
   The \NR\ prediction fails to recover in detail the diffusion structures observed numerically in Fig.~\ref{fig:dFdt_NBODY_LCosI}.}
   \label{fig:dFdt_NR_LCosI}
 \end{figure*}

In this appendix, we follow the same approach
as in section~\ref{sec:Secular_cosI},
and investigate relaxation in ${ (L , \cos I) }$.
We refer to Appendix~\ref{app:FP_2d_LcosItrue}
for the derivation of the relevant 2D \FP\ equation.
In Fig.~\ref{fig:dFdt_NBODY_LCosI},
we illustrate the $N$-body measurements
while Fig.~\ref{fig:dFdt_NR_LCosI} presents the associated \NR\ predictions.

From these two figures, we can globally make
the same observations as in section~\ref{sec:Secular_cosI},
which considered diffusion in ${ (\Jr , \cos I) }$.
Indeed, in rotating clusters, diffusion reshuffles orbits
towards smoother distributions of inclinations.
Nevertheless, the \NR\ prediction fails at predicting the exact structures
measured in $N$-body simulations.
In particular, the \NR\ theory predicts a relaxation rate
varying weakly with $\cos I$ (for a given sign of $\cos I$).
This differs from the $N$-body simulations,
where the relaxation rate decreases away from ${\cos I\!=\!0}$.

\section{Impact of discontinuities}
\label{app:impact_disct}

\begin{figure} 
    \centering
   \includegraphics[width=0.45 \textwidth]{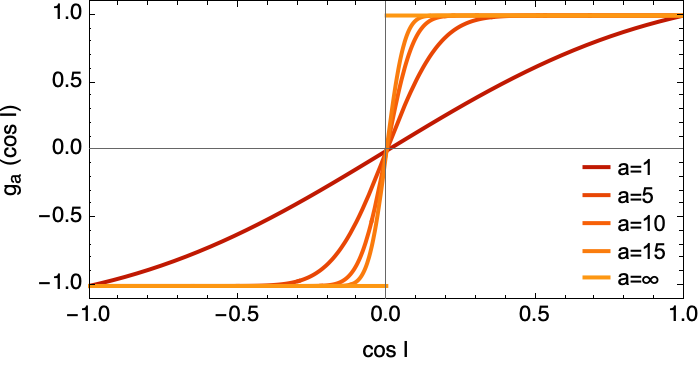}
   \caption{Family of functions $g_{a}(\cos I)$ (equation~\ref{eq:ga}) for various $a$.
   For $a \!\rightarrow\! 0$, the function approaches identity,
   while for ${a\! \rightarrow \! \infty}$, it approaches the sign function.
   Varying $a$ allows us to investigate the impact
   of the \LBD\ discontinuity in equation~\eqref{eq:LBD}.
   }
   \label{fig:g_a}
 \end{figure}

In this appendix, we investigate in more detail
the impact of the discontinuity at ${ \cos I \!=\! 0 }$,
introduced by the \LBD\ (equation~\ref{eq:LBD}).
To do so, we follow the same approach as in section~{2.3} of~\citet{Rozier2019} 
and consider rotating \DFs\ of the form
\begin{equation}
\Frot(\Jr,L,\Lz) = \Ftot(\Jr,L) (1+\alpha g[\Lz/L]),
\end{equation}
where $g[\cos I]$ is an odd function with ${ g (1) \!=\! 1 }$. 
The  \LBD\ corresponds to ${g\!=\!\sgn}$.
To approximate smoothly the \LBD\@,
we consider the sequence of functions
\begin{equation}
\label{eq:ga}
g_a(x) =  \erf(a x) / \erf(a) .
\end{equation}
As illustrated in Fig.~\ref{fig:g_a},
this ensures that ${g_0(x)\!=\!x}$ and ${g_{\infty}(x)\!=\!\sgn(x)}$.  

\subsection{The $\cos I$ coordinate}

\begin{figure} 
    \centering
   \includegraphics[width=0.45 \textwidth]{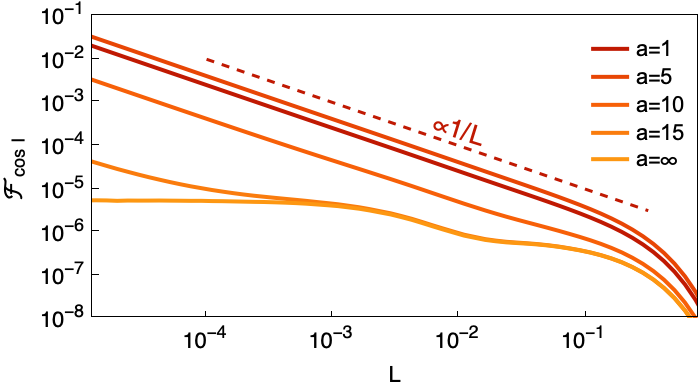}
   \caption{ Diffusion flux along $\cos I$ of the 3D \FP\ equation
   in ${ (\Jr , L , \cos I) }$, for ${\Jr\!=\!0.1}$ and ${\cos I \!=\!0.2}$,
   as a function of $L$ and for various smooth functions $g_a$ (equation~\ref{eq:ga}). Here, we consider an isotropic cluster with rotation parameter ${\alpha\!=\!0.25}$. 
   For any smooth $g_a$, the flux diverges like ${1/L}$ for ${ L \!\to\! 0 }$.
 As ${ a \!\to\! \infty }$,
that is, as $g_a$ tends to the $\sgn$ function,
 the flux converges to the \LBD\ flux pointwise,
which exhibits no divergence.
   }
   \label{fig:FluxCosI_a}
 \end{figure}

To probe the possible presence of discontinuities and singularities,
let us first compute the $\cos I$ component of the 3D flux
in ${ (\Jr , L , \cos I) }$, as introduced in appendix~\ref{app:convert_Lz_to_cosI}.
The dependence of this flux with respect to $a$ is illustrated in Fig.~\ref{fig:FluxCosI_a}.

In this figure, for any smooth $g_{a}$, we observe a $1/L$ divergence
of the flux as ${ L \!\to\! 0 }$.
In a nutshell, when working with ${ \cos I }$,
we suffer from a coordinate singularity,
and the \NR\ prediction cannot be applied to a \DF\ with a smooth rotation function.
Yet, when ${a \!\rightarrow \! \infty}$, the flux converges towards the LBD flux pointwise,
and this flux does not diverge for ${ L \!\to\! 0 }$.
Phrased differently, in the particular case of the \LBD\ sign function,
we can make a meaningful and well-posed \NR\ prediction
for the diffusion in ${ \cos I }$.
In practice, this vanishing of the divergence stems
from the cancellation of the derivative of ${ \sgn (\cos I) }$ everywhere
(except for ${\cos I\!=\!0}$).
Such a property is not the norm
for smooth arbitrary rotation functions, ${ g (\cos I) }$.
In that case, the $1/L^2$ singularities visible in equations~\eqref{eq:change_var_Lz_cosI}
do not combine into an integrable quantity. 

\begin{figure} 
    \centering
   \includegraphics[width=0.45 \textwidth]{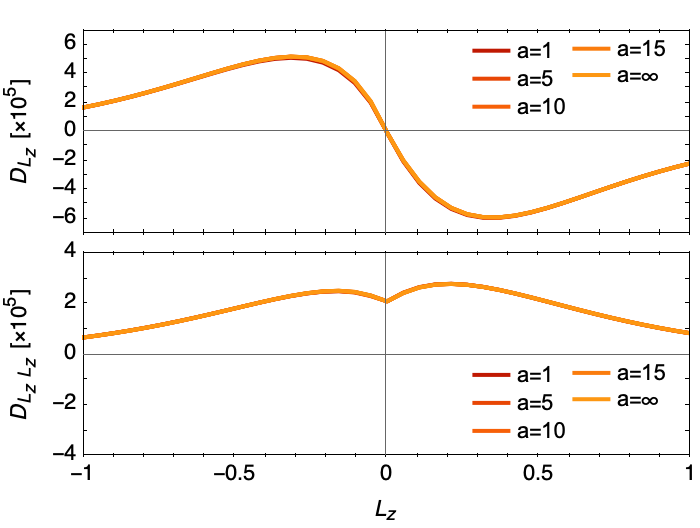}
   \caption{ One-dimensional diffusion coefficients  $\overline{D}_{\Lz}$ (top panel) and $\overline{D}_{\Lz\Lz}$ (bottom panel) of equations~\eqref{eq:DLz_DLzLz}, computed for the rotating isotropic clusters with ${\alpha \!=\! 0.25}$,
   and with the smoothing parameters ${a \!=\! 1,5,10,15,\infty}$, as defined in equation~\eqref{eq:ga}.
   These coefficients are well-defined for all values of $\Lz$.
   In addition, they only (very) weakly depend on $a$,
   with relative differences of order 1--5\%.
      }
   \label{fig:DLz_DLzLz}
 \end{figure}

To further stress that this divergence originate
from coordinate singularities,
let us now produce \NR\ predictions
by describing relaxation in ${ (\Jr , L , \Lz) }$
rather than in ${ (\Jr , L , \cos I) }$.
To do so, we define the \DF\ in $\Lz$ as
\begin{equation}
F(\Lz)= \int_{0}^{+ \infty} \!\!\!\! \rd \Jr \, \!\! \int_{|\Lz|}^{+ \infty} \!\!\!\! \rd L\, \Frot(\Jr,L,\Lz).
\end{equation}
Integrating the 3D \FP\ equation~\eqref{eq:def_FP} over $\Jr$ and $L$ yields 
\begin{equation}
\frac{\p F (\Lz)}{\p t} = -\frac{\p }{\p \Lz}\big(\overline{D}_{\Lz} F[\Lz]\big) + \frac{1}{2} \frac{\p^2}{\p \Lz^2}\big( \overline{D}_{\Lz\Lz} F[\Lz]\big) .
\label{eq:FP_Lz}
\end{equation}
Here, the 1D diffusion coefficients in $\Lz$ are given by
\begin{subequations}
\label{eq:DLz_DLzLz}
\begin{align}
\overline{D}_{\Lz} &= \!\! \int_{0}^{+ \infty} \!\!\!\! \rd \Jr \, \!\! \int_{|\Lz|}^{+ \infty} \!\!\!\! \rd L\,D_{\Lz} \Frot(\Jr,L\,|\,\Lz),\\
\overline{D}_{\Lz\Lz} &= \!\! \int_{0}^{+ \infty} \rd \Jr \, \!\! \int_{|\Lz|}^{+ \infty} \!\!\!\! \rd L\,D_{\Lz\Lz} \Frot(\Jr,L\,|\,\Lz),
\end{align}
\end{subequations}
with ${\Frot(\Jr,L\,|\,\Lz)\!=\!\Frot(\Jr,L,\Lz)/F(\Lz)}$
standing for the \DF\ in $(\Jr,L)$ given $\Lz$,
and normalised to unity.
Importantly, for ${ \Lz \!\neq\! 0 }$,
the coefficients $\overline{D}_{\Lz}$ and $\overline{D}_{\Lz\Lz}$
are well defined whatever the considered rotating \DF.
In addition, they both converge to some finite values as ${ \Lz \!\to\! 0 }$,
as illustrated in Fig.~\ref{fig:DLz_DLzLz}.
The absence of any divergence here emphasizes
that the divergence observed in Fig.~\ref{fig:FluxCosI_a}
stems from a coordinate singularity associated with ${ \cos I }$.

\subsection{Impact on $N$-body measurements}
\label{app:lbd_v_erf}

\begin{figure} 
    \centering
   \includegraphics[width=0.45 \textwidth]{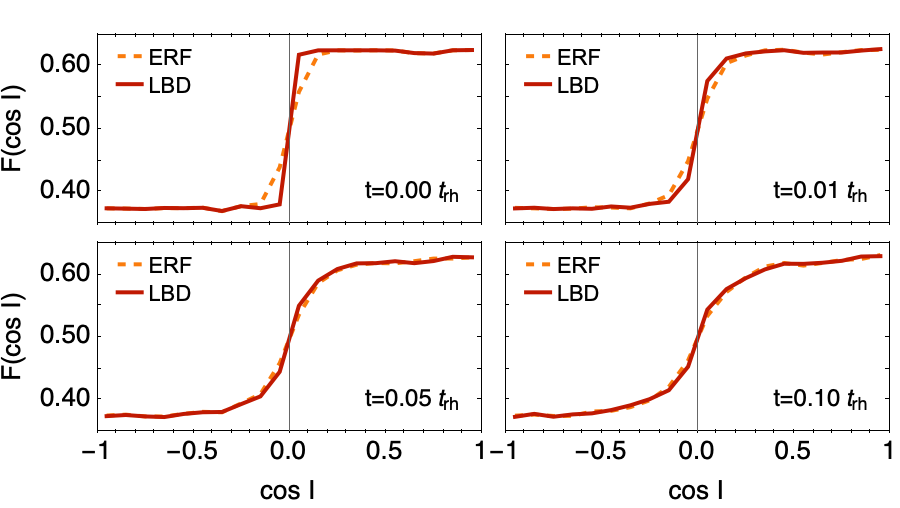}
   \caption{  Comparison between two rotating, isotropic Plummer clusters with parameters $(q,\alpha)\!=\!(0,0.25)$ and $N\!=\!10^5$ stars. The discontinuous cluster follows the LBD parametrization (equation~\ref{eq:LBD_cosI}), while the smooth cluster follow an $\erf$ approximation, as defined in eq.~\eqref{eq:ga}. Both measurements have been ensemble-averaged over 10 realizations. Beyond the initial dilution of the discontinuity, we observe that the initially discontinuous DF joins the smooth DF, after which both systems appear to follow the same evolution. 
      }
   \label{fig:ldb_erf}
 \end{figure}

To estimate the impact of the discontinuity on relaxation, we used $N$-body
simulations in Fig.~\ref{fig:ldb_erf} to compare the early relaxation
of the discontinuous LBD distribution (equation~\ref{eq:LBD})
with its smooth approximation (equation~\ref{eq:ga} with ${ a \!=\! 10 }$).
Reassuringly, we observe a fast dilution of the discontinuity, with the DF quickly resembling its smooth approximation. Afterwards, no significant difference is observed between the two systems.

\end{appendix}

\end{document}